\documentclass[10pt]{article}
\usepackage[ansinew]{inputenc}
\usepackage{fancyhdr}

\usepackage{sectsty} 
\usepackage[normalem]{ulem}
\allsectionsfont{\large}
\subsectionfont{\normalsize}

\usepackage{graphicx}
\usepackage{color}
\usepackage[colorlinks=true, citecolor=blue, linkcolor=blue, urlcolor=blue]{hyperref}
\usepackage[active,new,noold,marker]{xrcs}
\usepackage[active]{srcltx} 
\usepackage{amssymb}
\usepackage{amsmath}
\usepackage{mathenv}
\title{\begin{flushright} \vspace{-70pt}\textit{\footnotesize Computer Science \& Engineering, \\ \vspace{-10pt} Phase: Theor. Proj.} \\
\textit{}\\\vspace{-20pt}
\textit{\footnotesize Tec. Rep. }\vspace{4pt}\footnotesize{(2007)}\footnotesize\textbf{\textit{ Vol.}}\textbf{ 1}, \emph{Rev. Ver.} 6, 1-33\\ \vspace{-1pt} \footnotesize Proj. No.TXU001347562 Ext. on 02 Oct 2007
\\ --------------------------------------------------------------- \end{flushright} \vspace{12pt} \Large \textbf {Logic, Design \& Organization $\mathrm{\textit{of}}$ PTVD-SHAM; A Parallel Time Varying \& Data Super-helical Access Memory}\\ }

\begin{document}

\definecolor{MyBrown}{rgb}{0.6,0.4,0}

\date{}
\maketitle
\begin{center}

 \vspace{-50pt}
\long\def\symbolfootnote[#1]#2{\begingroup%
\def\thefootnote{\fnsymbol{footnote}}\footnote[#1]{#2}\endgroup}

\textbf{}

By Philip Baback Alipour \(^{1}\,^{,}\, ^{2}\,^{,}\) \noindent\symbolfootnote[1]{Author for correspondence (\htmladdnormallink {\textcolor{blue}{philipbaback\_orbsix@msn.com}}{mailto:philipbaback_orbsix@msn.com}).} $^{,}$ \noindent\symbolfootnote[2]{The original version of this paper was created in MS-Word software program, where the current version is reconstructed from the original paper in LaTeX editing environment for public review. I thank those anonymous administrators at ArXiv.org, rectifying file conversion problems on the previous versions of the paper since the 1st submission on 9 July 2007.}
\end{center}

\begin{center}
\textit{1-Category of Computer Sciences, Laboratory of Systems Technology, Research, Design \& Development,}

\textit{ Elm Tree Farm, Wallingfen Lane, Newport, Brough, HU15 1RF, UK}

\smallskip
\textit{2- Computer Science \& Engineering Departments, University of Hull, Cottingham Road,           Hull Campus, East Yorkshire, HU6 7RX, UK}

\end{center}

\noindent \textbf{Abstract. }\small{This paper encompasses a super helical memory system's design, `Boolean logic \& image-logic' as a theoretical concept of an invention-model to `store time-data' in terms of anticipating the best memory location ever for data/time. A \textit{waterfall effect} is deemed to assist the process of potential-difference output-switch into diverse logic states in quantum dot computational methods via utilizing coiled carbon nanotubes (CCNTs) and carbon nanotube field effect transistors (CNFETs). A `quantum confinement' is thus derived for a flow of particles in a categorized quantum well substrate with a normalized capacitance rectifying high \textbf{B}-field flux into electromagnetic induction. Multi-access of coherent sequences of `qubit addressing' is gained in any magnitude as pre-defined for the orientation of array displacement. Briefly, Gaussian curvature of $k<0$ is debated in aim of specifying the 2D electron gas characteristics in scenarios where data is stored in short intervals versus long ones e.g., when $k'>\left(k<0\right)$ for greater CCNT diameters, space-time continuum is folded by chance for the particle. This benefits from Maxwell-Lorentz theory in Minkowski's space-time viewpoint alike to crystal oscillators for precise data timing purposes and radar systems e.g., time varying self-clocking devices in diverse geographic locations. This application could also be optional for data depository versus extraction in the best supercomputer system's locations, autonomously. For best performance in minimizing current limiting mechanism including electromigration, a multilevel metallization and implant process forming elevated sources/drains for the circuit's staircase pyramidal construction, is discussed accordingly.}

\begin{center}
\vspace{8pt}
\smallskip
\end{center}

\section{Principles and introduction}
\label{section1}
\vspace{1pt}
\small Possessing the fact that multiple qubits can exhibit quantum entanglement (see e.g., qubit entanglement~\cite{27-Wiki}), a solution to store the entangled Bell states of electrons from 2DEG system, moving free into the two-dimensional plane, now to some complex plane's depository vector system defining storable qubit(s), is thus emerged in practice.

Furthermore, binary values by binary variable $\mathrm{bit}_{i} $, as input signal $\forall A_{in} \in \left\{0,1\right\}$ by ratio to qubit variable as input signal $B_{in} $, are introduced via bit-frequency
\[
\nu _{bit}  \equiv \frac{{A_{in} }}{{B_{in} t}}\left| {\left. {{\mathop{\rm for}\nolimits} \; A_{in} \left\{ \begin{array}{l}
 0 \buildrel \wedge \over = {\mathop{\rm L}\nolimits} _{ \pm V}  \\
 1 \buildrel \wedge \over = {\mathop{\rm H}\nolimits} _{ \pm V}  \\
 \end{array} \right.} \right\}{\mathop{\rm for}\nolimits}\; {\rm  }B_{in} :\left| \Psi  \right\rangle  = \prod\limits_{m = 1}^n {\left| \Psi  \right\rangle _m } } \right.,{\rm  }m = 1,2,...,n, \:\: \ (1.1)
\]

\noindent which is in the memory system's product, hereby of importance to generate quantum efficiency for hot-electron spectra. With response to mathematical relations in defining the characteristics of the storage layers of the memory system (subsequent relation), relation (1.1), distributes into future relations (5.1) to (5.5), explaining events related to low-voltage differential signaling (LVDS) e.g.,~\cite{38-Texas Inst.}. Specifically, these voltages are computed between components of the memory's staircase levels in terms of $V_{in} $, addition to stored charges possessing a quantized voltage $\ge k\:\mathrm{mV}$ amongst memory traps $\mathrm{for}{\rm \; }B_{in} $ defined from wavefunction ${\left| \psi  \right\rangle} $'s Bell states.

\indent In possession of the previous relation, simplifying Bell states' relations concerning collectable data points as quantum dots embedded in a 2DEG layer of the high electron mobility transistor (HEMT; for higher state of carrier's mobility) and metal-oxide semiconductor field-effect transistor (MOSFET technology), did not require the study of spin conditions. The requirement to examine the four well-known Bell states' scenarios during the course of entanglement in the 3D space where CNT coils remain, was also eliminated. In fact, from the point of particle confinement, a depository vector emergence by some arbitrary angle for a storable qubit in the storage film, to the point of storing Boolean and qubit data as the lower layers of the semiconductor's gate electrode, is studied theoretically. In addition, the present chip design's architecture appreciates Choi \textit{et al.} of~\cite{06-Choi}'s `MOSFET technology utilizing CNTs', to some extent in application's design basics thus not entirely covering the final design outlook itself.

\indent The latter and former paragraphs are preferred as the main objective of the fabrication process rather than studying the probabilistic nature of entangled N-tuplets between the remaining gates in a staircase manner. That is, \textit{quantum confinement against entanglement Fermi-level regime, where the Fermi energy is ought to be determined where the electron concentration and effective mass}~\cite{27-Wiki} \textit{meet}.

In this case, it is imperative to assess the mass confined into the depository's vector bands of the memory for an observable quantum efficiency quantity within the memory cell's domain (e.g., compare with~\cite{05-Chen}). In contrast, here, `quantum efficiency' is not of `the number of collected electrons over the number of incident photons' for a solar cell defining its level of responsivity via factor of absorption coefficient. In fact, this quantity is studied by the perception that supports resonant cavity layers to the actual memory cell layers of the chip in terms of
\vspace{-10pt}
\begin{flushleft}
$$\eta _{{\bf f}} =\frac{output_{{\bf f}} }{input_{{\bf f}} }= \frac{{\rm \; number\; of\; electrons\; collected}}{{\rm storable\; number\; of\; ballistic\; hot\; electrons\; in\; entanglement\; \; }}  \; \; \; \; \; \; \; \; \; \; \; \; \; \; \; \; \; \; \; \; \; \; \; \; \; \; \; \; \; \; \; \; \; \; \; \; \; \; \;$$ $$= \frac {N_{e}}{ N_{e\stackrel{L_{V} ,{\rm \; }H_{V} }{\longrightarrow}{\left| \Psi  \right\rangle} }}. \; \; \;  \; \; \; \; \; \; \; \; \; \; \; \; \; \; \; \; \; \; \; \; \; \; \; \; \; \; \; \; \; \; \; \; \; \; \; \; \; \; \; \; \; \; \; \; \; \; \; \; \; \; \; \; \; \; \; \; \; \; \; \; \; \; \; \; \; \; \; \; \; \; \; \; \; \; \; \; \; \; \; \; \; \; \; \;  \; \; \; \; \; \; \; \; \; \; \; \; \; (1.2)$$
\end{flushleft}

\noindent This relation is of fermion type; both quanta input and output paradoxically are of fermions. Thus, the term `fermionic quantum efficiency' stands pertinent for hot electrons or hot carriers denoting the principle behaviour of quasi-Fermi characteristics upon the incoming hot electrons from 2DEG staircase layers in entanglement.

The latter carrier characteristics is due to the grouping result of poles: $\alpha $, $\beta $ in Fig.$\:2.1$, \S\ref{section2}. This effect comes from CCNTs that subjugate mobile carriers to act directional in favour of the entangled wavevector approaching storage cells for all incoming electrons from the 2DEG system to 3D-space, subsequently, committed to a 2D-quantum capacitor of the memory cell. The entangled wavevector is hence derived via, ${\rm {\mathfrak E}}\equiv \left(e\sqrt{-1e_{2} e'_{2} } \right)\dot{t}$, elicited from relations (4.1) to (4.11), \S\ref{section4}, in satisfaction of the context of Hyp.$\,$4.1. In continue,
\vspace{-18pt}
\begin{flushleft}
\[\prod \limits _{}^{}E_{e\stackrel{L_{V} ,{\rm \; }H_{V} }{\longrightarrow}{\left| \Psi  \right\rangle} }  =\frac{1}{nE_{{\left| \Psi  \right\rangle} } } \mathop{{\rm \int }}\nolimits_{d}^{^{} } {\rm \; }F_{e} \cdot {\rm \; }\mathrm{d}d_{xy,z} =\frac{E_{e,z} \times E_{{\rm {\mathfrak E}}} +E_{e,xy} \times E_{{\rm {\mathfrak E}}} }{E_{{\left| \Psi  \right\rangle} } } {\rm ,\; and,\; \; \; \; \; \; \; \; \; \; \; \; (1.3)}\]
$$\because E_{e,xy} \stackrel{E_{{\left| \Psi  \right\rangle} } }{\longrightarrow}\forall E_{{\rm {\mathfrak E}}} \in \left. k_{C} Q\sqrt{QQ'} \left(2\sqrt{xy} \: \mathbf{o}A_{x_{i} y_{i} } \right)^{-1} \right|k_{C} =\frac{1}{4\pi \varepsilon _{\circ } } \approx {\rm 8.988\times 10}^{{\rm 9}} \rm{Nm^{{\rm 2}} C^{-2}} ,$$
thus,\vspace{-6pt}
\[=\left(k_{C} \left|{\rm {\mathfrak E}}\right|\left\{\frac{\hbar ^{2} k_{x}^{2} }{2m_{e} } +\frac{\hbar ^{2} k_{y}^{2} }{2m_{e} } +\frac{\hbar ^{2} k_{z}^{2} }{2m_{e} } \right\}\right)^{\frac{1}{2} } =\sqrt{{\rm \; }\frac{k_{C} \left|{\rm {\mathfrak E}}\right|\hbar ^{2} }{4d_{xy,z} m_{e} } } =\sqrt{{\rm \; }\frac{k_{C} \left|{\rm {\mathfrak E}}\right|\hbar ^{2} }{4{\rm {\mathcal V}}_{xyz} m_{e} } } \ge 0{\rm \; }\mathrm{J}\]
\end{flushleft}

                         , $\left|{\rm {\mathfrak E}}\right|^{2} \equiv e^{2} e_{1} e_{2} \dot{t}^{2} $,  $k=\left(k_{x} ,k_{y} ,k_{z} \right)$ ,  $d\equiv x{\rm \: }\mathbf{o}k,{\rm \; }y{\rm \: }\mathbf{o}k,{\rm \; }z{\rm \: }\mathbf{o} k , \; \; \; \;\; \; \; \; \; \; \; \; \; \; \; \; \; \; \;\; \; \; \; \; \; \; \; \; \; \; \; \;(1.4)$
\vspace{2pt}

\noindent wwhere $E_{{\left| \Psi  \right\rangle} } $, represents the probable energy expectation on wavefunction, ${\left| \psi  \right\rangle} $'s Bell states product from (1.1), for the amount of work done in spatial dimensions $d_{xy} $ satisfying $E_{{\rm {\mathfrak E}}} $ as the energy of the entangled wavevector, ${\rm {\mathfrak E}}$. Dimension $d_{z} $ in return, satisfies $E_{e,z} $ as the energy of the 2DEG or in general, the energy of D-dimensional, $g\left(E_{e} \right)\sim E_{e}^{\left(D-2\right)2^{-1} } $, from the point of quantum wells to the point of charge's storage trap(s), correspondingly. In addition, $\mathbf{k}$ for energy $E_{e,z} $, is in regard to Leadley of~\cite{22-Leadley}'s wavevector definition used to describe a free particle of mass $m$ when it's in confinement, however, not being in entanglement, e.g., the choice of \textit{z}-direction in the 2D system. Entangled energy $E_{{\rm {\mathfrak E}}} $, in the energy product is of scalar form, thus the magnitude of energy is of interest and not its direction, since the direction is already exhibited by directions of $E_{{\left| \Psi  \right\rangle} } $ in the denominator's distribution area via implication, $E_{e,xy} \stackrel{E_{{\left| \Psi  \right\rangle} } }{\longrightarrow}E_{{\rm {\mathfrak E}}} $. In other words, the more energy distribution of $E_{e} $ in favour of direction $z$ against $xy$, the more discrete the energy product and thus, partitioned internally. Now, having from relation (2.2) in~\cite{10-Delaney},

\vspace{8pt}

$H_{C} \psi =E_{C} \psi ,\:\:\:\:\:\:\:\:\:\:\:\:\:\:\:\:\:\:\:\:\:\:\:\:\:\:\:\:\:\:\:\:\:\:\:\:\:\:\:\:\:\:\:\:\:\:\:\:\:\:\:\:\:\:\:\:\:\:\:
\:\:\:\:\:\:\:\:\:\:\:\:\:\:\:\:\:\:\:\:\:\:\:\:\:\:\:\:\:\:\:\:\:\:\:\:\:\:\:\:\:\:\:\:\:\:\:\:\:\:\:\:\:\:\:\:\:\:\:\:\:\:\:\:(1.5)$                                                                                       \vspace{8pt}

\noindent and (5.13) from~\cite{10-Delaney}, whilst benefiting from the current report'relations (1.3) to (1.5), thence,
\vspace{-6pt}
\[\therefore E_{C} \psi =\frac{EE_{e} }{\prod E_{e\stackrel{L_{V} ,{\rm \; }H_{V} }{\longrightarrow}{\left| \Psi  \right\rangle} }  } \prod\limits_{i = 1}^n {\phi _i } \; , \;  i=1,2,...,n \: . \:\:\:\:\:\:\:\:\:\:\:\:\:\:\:\:\:\:\:\:\:\:\:\:\:\:\:\:\:\:\:\:\:\:\:\:\:\:\:\:\:\:\:\:\:\:\:\:\:\:\:\:\:\:\:\:\:\:\:\:\:\:\:\: \; \; \; \; \; (1.6)\]
\vspace{-6pt}

\noindent Let $E_{e} $, serve the classical energy expectation on electrons representing circuit component's classical Boolean logic $E$, which is the non-local energy expectation from the circuit. This non-local energy expectation must be the remaining energy across the circuit sever from $E_{e} $ and $E_{e\stackrel{L_{V} ,{\rm \; }H_{V} }{\longrightarrow}{\left| \Psi  \right\rangle} } $, but not the `local energy' which is the energy associated with the regions outside of the sub-system and thus their interactions with the sub-system (see relation 5.15 from the concept of local energy density in~\cite{10-Delaney}).

By means of which, the levels of degeneracy principle of energy-states from mobile carriers including their probable superposition to storable time and data space locations are discussed by definition. This definition should be relevant to the nature of circuit's capacitance, i.e., degeneration of eignstates corresponding to identical eignvalues of the Hamiltonian in aim of maintaining a symmetry between `electric potential energy states' (observe relation 1.6) and `electric flux density'. The degeneracy behaviour in principle, appears from Maxwell's equations in this case, is by clarifying bit frequency $\nu _{bit} $'s physical property and characteristics. Symbol $\phi _{i} $, is one-electron wavefunctions and $k_{C} $, in singled-one-electron product of (1.4), is the electrostatic constant from Coulomb's law, whereas without the consideration of $e_{1} e_{2} \dot{t}$'s relative existence, the remaining component $k_{C} ed^{-1} $, is therefore measured in volts.

Perceivably, the atomic units, $m_{e} $, $\hbar $ and $e$, for the electronic circuit at time $t=0$, represent a closed loop of Hamiltonian type denoting the time evolution of quantum states for the entire circuit once the current flows, hence computing the total energy of the system. It is also perceived that in this case, all corresponding equations are of ratio time-dependent Schrödinger's equation (not time-dependent) via time factor $\dot{t}$ of ${\rm {\mathfrak E}}$. This type of equation is governed to be conserved for sufficient capacity, consequently maintaining a time-independent Schrödinger equation expressing a steady-state similar to the indication made on `quantum sub-systems as circuit blocks', discussed by Delaney \& Greer in~\cite{10-Delaney}. Once the shift of pure state at time $t\ge 0$ occurs at some closed-loop state of eignvalues via depository vectors on the memory components, here, the waterfall input/output observable operators, the system's performance as the circuit's power supply is deemed to be not in a pure state any longer. This shift from pure state to a new state occurs, once the fuzziness of 2DEG system and CCNTs comes into practice. In such scenarios, the correlation of inputs and outputs for the involved subsystems could be expressed through the expectation values:

\vspace{-2pt}
\begin{flushleft}
$\left\langle \hat{i}\hat{o}\right\rangle \ne \left\langle \hat{i}\right\rangle \left\langle \hat{o}\right\rangle $.    $\:\:\:\:\:\:\:\:\:\:\:\:\:\:\:\:\:\:\:\:\:\:\:\:\:\:\:\:\:\:\:\:\:\:\:\:\:\:\:\:\:\:\:\:\:\:\:\:\:\:\:\:\:\:\:\:\:\:
\:\:\:\:\:\:\:\:\:\:\:\:\:\:\:\:\:\:\:\:\:\:\:\:\:\:\:\:\:\:\:\:\:\:\:\:\:\:\:\:\:\:\:\:\:\:\:\:\:\:\:\:\:\:\:\:\:\:\:
\:\:\:\:\:\:\:\:\:\:\:\:\:(1.7)$  \textit{   }
\end{flushleft}
\vspace{-2pt}

\noindent Expressing that, `the inputs and outputs are specified in terms of subsystem observables, or operator expectation values, and these quantities must correlate through a density matrix'~\cite{10-Delaney}, wherein this paper, for the ratio-time-dependent state, we merge the concept into
\smallskip
\begin{flushleft}
$\left\langle \hat{i}\hat{o}\right\rangle \xrightarrow[{{\rm \int }\mathrm{d}\tau }]{{\rm \int }\mathrm{d}t} \left\langle \hat{i}\hat{o}\right\rangle =\left\{\left\langle \hat{i}\right\rangle t\frac{\left\langle \hat{o}_{i} \right\rangle }{\tau _{\Sigma } } \right. =\left\langle \hat{o}\left(\hat{i}\left(\hat{t}\right)\right)\right\rangle =f\left(\hat{o}\right)=\left\langle \hat{o}\hat{\tau }_{0} \right\rangle $ , $\tau _{\Sigma } \equiv \tau _{0} =1\left(t_{1} \left|t_{2} \right. \right),1\left(t_{2} \left|t_{1} \right. \right)$ , $\:\:\:\:\:\:\:\:\:\:\:\:\:\:\:\:\:\:\:\:\:\:\:\:\:\:\:\:\:\:\:\:\:\:\:\:\:\:\:\:\:\:\:\:\:\:\:\:\:\:\:\:\:\:\:\:\:\:
\:\:\:\:\:\:\:\:\:\:\:\:\:\:\:\:\:\:\:\:\:\:\:\:\:\:\:\:\:\:\:\:\:\:\:\:\:\:\:\:\:\:\:\:\:\:\:\:\:\:\:\:\:\:\:\:\:\:\:
\:\:\:\:\:\:\:\:\:\:\:\:\:(1.8)$

$\therefore \left\langle \hat{i}\hat{o}\right\rangle =\left\langle \hat{o}\hat{\tau }_{0} \right\rangle \left\langle \hat{i}\hat{t}\right\rangle $.                                                                                  $\:\:\:\:\:\:\:\:\:\:\:\:\:\:\:\:\:\:\:\:\:\:\:\:\:\:\:\:\:\:\:\:\:\:\:\:\:\:\:\:\:\:\:\:\:\:\:\:\:\:\:\:\:\:\:\:\:\:
\:\:\:\:\:\:\:\:\:\:\:\:\:\:\:\:\:\:\:\:\:\:\:\:\:\:\:\:\:\:\:\:\:\:\:\:\:\:\:\:\:\:\:\:\:\:\:\:\:\:\:\:\:\:\:\:\:\:\:
\:\:\:\:\:\:\:\:\:\:\:\:\:(1.9)$
\end{flushleft}
\vspace{-2pt}
Let time $t$, be empirically dedicated to operator expectation input value $\hat{i}$, and asymptotically, the stored proper time $\tau _{\Sigma } $ dedicated to operator expectation output value $\hat{o}$. By composition, the interdependency with inputs and their storable proper time including stationary time (or, time at rest) is carried out via, $\hat{o}\left(\hat{i}\left(t\right)\right)$, once the time factor integrals are implied to operator expectation values, $\left\langle \hat{i}\hat{o}\right\rangle $. Now, by realizing the time dilation principle~\cite{25-Nave,{41-Wright}}, and Einstein's relativity~\cite{14-Einstein et al.}, one could herein deduce, \\

\noindent \textbf{Theorem 1.1.} \textit{It is supposed that, storable proper time to be the time given to a particle as its time property on its worldline connecting geometric points of field lines of (4.1) with $e^{2} e_{1} e_{2} \dot{t}^{2} $ of (4.2), \S\ref{section4}. The following possessions indicate the conditions of this theorem: } \\\vspace{-12pt}

\begin{enumerate}
\item [{(1)}]\textit{Let $e^{2} e_{1} e_{2} \dot{t}^{2} $ possess stationary time $t_{1} $ and $t_{2} $ of $\mathrm{d}t\leftrightarrow \tau $ from (4.3) for its unaccelerated worldline with respect to accelerated worldline performed by the involved particle.}
\vspace{-1pt}
\item [{(2)}]\textit{The accelerated worldline in possession (1), could be of Gaussian curvature of $k<0$ representing for $\hat{i}$'s data, which is being stored in short intervals versus long intervals when the mobile particle's worldline super-symmetry performs, $k'>\left(k<0\right)$, in the context of space-time simultaneity from special relativity.} \\
\vspace{-12pt}
\item [{(3)}]\textit{In possession, these states of curve in (2), could lead to homogeneity of time quantities $t_{1} \left|t_{2} \right. $ and $t_{2} \left|t_{1} \right. $ despite of their highly symmetric relationships (4.2) and (4.3) by representing time factor paradox, $\dot{t}^{2} \ne \dot{t}^{2} $.}
\end{enumerate}

\noindent \textbf{Proof 1.1.} \textit{In essence, commences with (4.2), \S\ref{section4}, followed by the subsequent logic, then rationalized by, (4.18) \& Hyp.$\,$4.1, thereby concludes with, Hyp.$\,$4.2.}\textit{}\\

\small Let the above-mentioned particle in Theorem 1.1, be a storable electron of charge, $e$, from special relativity (SR). The stored proper time for the inputted data is thereby emerged for its future time in entanglement via `time barrier' symmetry $\left(t_{1} \left|t_{2} \right. \right)$ or $\left(t_{2} \left|t_{1} \right. \right)$, such that, data displacement from one memory location to another throughout the course of time is plausible via the circuit itself as, `internal data partitioning'. The so-called `one memory location' is denoted by a `$1$' which is multiplied by the time symmetry as a unit defining `data displacement'.

Clearly, there is neither input/output (I/O) external data-partitioning nor I/O union compositions, whereas the given confined `quantum timing frame' is prior to data type and bus subsystem's way of allocating and program compilation process. This time frame condition is discussed broadly in \S\ref{section4}. The chip's dependence upon the I/Os, this time frame establishment therefore encloses operator expectation values, $\left\langle \hat{i}\hat{o}\right\rangle $ and $\left\langle \hat{i}\right\rangle \left\langle \hat{o}\right\rangle $ into a `satisfied correlated I/O expectation value', $\left\langle \hat{i}\hat{t}\right\rangle $ and $\left\langle \hat{o}\hat{\tau }_{0} \right\rangle $, respectively. These `correlated I/O expectation values' are for those interconnected gates with multiple inputs switching simultaneously, which is due to the time and data loop's characteristics, consistent to the following argument's paradox:

\vspace{6pt}

\noindent --------------- Contradictorily, as we confront in \S\ref{section4}, `time' for these subsystems behaves in accordance with Russell's Paradox, just as generalized in the calculation made on (1.8) and (1.9). So, the partitioning of a system from a quantum mechanical viewpoint is a resolved point, since the time problem itself is storable in addition to `read and write data input' process. Hence, perpetuating the steady-state in a loop of subsystems' inputs in correlation to outputs as `readable data' is proceeded, no matter the degree of freedom defining the role of density matrices, the data in-and-out process is partitioned periodically. That is, partitioning and storing proper time according to future closure relation, (4.14b), \S\ref{section4}, specifying, $\tau _{\Sigma } \approx 0.7071\left|\dot{t}_{0} \right|$ whilst time, $t$, is ticking as $t\ge 0$ for the next data write time phase, \textit{$WR.\phi _{1} $}, and read time phase, \textit{$RD.\phi _{1} $}, operation. ---------------
\vspace{6pt}

Recalling (1.4), we have used the Pythagoras theorem reasoning that, $E_{{\left| \Psi  \right\rangle} } $, as hypotenuse according to the initial theorem and, $d$, later known as radial time distance, $r$, from (4.15), \S\ref{section4}. This is the distance recalled for the geometry of revolution in favour of `depository vector', $\Sigma $,  in the context of Hyp.$\,$4.1. Thus, the entangled wavevector ${\rm {\mathfrak E}}$ in all product relationships, should also possess a complex number value representing a complex plane in the space and path of electron's transportation. Therefore, this type of quantum efficiency describing products of (1.2) through (1.4), fulfils `Image Logic' of the storing data, where the necessity of installing catalysts and absorbers across the chip is considered to be vital during chip's fabrication.

In practice, since computer memories require a capacitor that preserves stored charges and a transistor serving as a switch in aim of write data or read data from the capacitor, possessing high transconductance is a must, where FETs in this case, HEMT (e.g.,~\cite{26-Okita}) and MOSFETs are deemed to be the most appropriate technology for this design. To this account, the `waterfall staircase system' (Fig.$\,$2.1, \S\ref{section2}), possesses the switching ability in its fundamental structure. All electrodes in the circuit plan must satisfy write time phase \textit{$WR.\phi _{1} $}, and read time phase \textit{$RD.\phi _{1} $}, when relevant p.d.'s (potential difference occurrences) are applied between the chip itself and thus its components here as:
\vspace{6pt}

\noindent --------------- At least, two CNFET inverters with unequal electrode length $L$ and equal width $W$ by ratio, $L:W$, indicating circuitry's electrical resistance, next to the chip's $\alpha $-pole; $\beta $-pole at $\delta $-pole in the middle. These inverters are connected to a pass transistor logic (PTL) via e.g., a CNFET transmission gate representing time phase $\phi _{2} $, to refresh all `symmetric and asymmetric stored data' including, `asynchronous or non-coordinated time (early, past and future)/`synchronous stored time' and `data storing event', across the waterfall system into memory cell for either pole (see relations 1.10, 1.11 and Fig.$\,$2.1 of \S\ref{section2})! For this post-mentioned logic, `time storage, time phase $\phi _{2} $-values as rise-decay of time phase', is in aim of verifying the process of refreshing time cycles on the $\alpha $-pole, $\beta $-pole and $\delta $-pole data feedback. This is done by PTL layout adjacent to the $\delta $-pole, which is also completed with circuitry's substrate growth and construction phase. ---------------
\vspace{6pt}

\noindent Reasonably, the nanodots in the porous film (Fig.$\,$2.1\emph{c}, \S\ref{section2}) in polarity, are positioned below the memory cell's insulating film layer in contact with the CNT. These nanodots are filled with a charge storage material which is one of silicon (Si) and silicon nitride (${\rm Si}_{3} {\rm N}_{4} $) claimed by Choi \textit{et al.} in~\cite{06-Choi}, thus storing charges for a long period of time, which still requires the determination of $\phi _{2} $ problem in electrical functions. This requirement is due to prioritizing the charges' autonomous repositioning of their space and time occupation across the waterfall system's components (discussed in \S\ref{section4} \& \S \ref{section5}).

In general, the porous film is made of aluminium oxide (${\rm Al}_{2} {\rm O}_{3} $). The CNT mentioned, is not of a pillar type CNT, and is of semiconducting type, where pillar-CNT by comparison should possess an ohmic electronic behaviour as metallic type unaffected by a gate voltage due to the role of CCNT adjacent to it. With compliance to time phase problems from VLSI basics by Pucknell \& Eshraghian of~\cite{30-Pucknell}, the time phase ratios involved in the poles distributed across the waterfall gate ohmic contacts for `ideal conditions', is given by
\vspace{-2pt}
\begin{flushleft}
\(\left\langle f\left(\phi _{1} \right)_{\left(\alpha ,\beta ,\delta \right)} \right\rangle =\alpha \frac{1}{n}\sum \limits _{i=1}^{n}\left(WR.\phi _{1} :RD.\phi _{1} \right)_{i} : {\rm \; }\beta \frac{1}{n} \sum \limits _{j=1}^{n}\left(WR.\phi _{1} :RD.\phi _{1} \right)_{j}\:\:\:\:\:\:\:\:\:\:\:\:\:\:\:\:\:\:\:\:\:\:\:\:\:\:\:\:\:\:\:\:\:\: \)
\vspace{-12pt}
\(=\delta \frac{1}{2} \sum \limits _{k=1}^{2}\left(WR.\phi _{1} :RD.\phi _{1} \right)_{k}  \:\:\:\:\:\:\:\:\:\:\:\:\:\:\:\:\:\:\:\:\:\:\:\:\:\:\:\:\:\:\:\:\:\:\:\:\:\:\:\:\:\:\:\:\:\:\:\:\:\:\:\:\:\:\:\:\:\:\:\:\:\:\:\:
\:\:\:\:\:\:\:\:\:\:\:\:\:\:\:\:\:\:\:\:\:\:\:\:\:\:\:\:\:\:\:\:\:\:\:\:\:\:\:\:\:\:\:\:\:\:\:\:\:\:\:\:\:\:\:\:\:\:\:
\:\:\:\:\:\:\:\:\:\:\:\:\:(1.10)\)
\end{flushleft}
\vspace{8pt}
whereby $n\ge 4$, and the dimensions of the transistors, grouping for these poles explaining sheet resistance, $R_{s} $, and measuring resistance, $R$, of thin films with uniform sheet thickness, $\theta $, including resistivity $\rho $, shall be,
\vspace{-1pt}
\begin{flushleft}$\left\{\forall L,W\in A\left|Z=L\cdot W^{-1} ,R=Z\cdot R_{s} ^{-1} ,R_{s} =\rho \theta ^{-1} \right. \right\}, \; \left\{\forall \square \in A\left|L=W,R=R_{s} \right. \right\},$\end{flushleft}

\noindent such that
\vspace{-2pt}
\begin{flushleft}
$\alpha ,\beta ,\delta .\sum \limits _{k=1}^{2}\left(WR.\phi _{1} :RD.\phi _{1} \right)_{k} \ne  \delta .\phi _{2} {\rm \; \; , \; \; }$$\alpha ,\beta ,\delta .\sum \limits _{k=1}^{2}\left(L:W\right)_{k} \ne \alpha '_{90^{\circ } } .\sum \limits _{l=1}^{2}\left(L:W\right)_{l}  ,\beta '_{90^{\circ } } $
\vspace{-6pt}
$ .\sum \limits _{m=1}^{2}\left(L:W\right)_{m}  $, and,
$\alpha ,\beta ,\delta .\sum \limits _{k=1}^{2}\left(L:W\right)_{k}  \approx \delta '.\left(L:W\right)\: . \ \ \ \ \ \ \ \ \ \ \ \ \ \ \ \ \ \  \ \ \ \ \ \ \  \ \ \ \ \ \ \ (1.11)$\end{flushleft}   \vspace{6pt}

\noindent Accordingly, cross-sectional area, $A$, in the recent predicates could be split into length $L$ and width $W$ describing ratio conditions of $Z$, henceforth, for any size of square, $\square $, giving resistance $R=R_{s} $. Architecturally, notations $\alpha , \; \beta $ and $\delta $ are just poles with a geometric unit value of `1' in Euclidian space dimensional chip's substrate-occupation, denoted by $\lambda ^{3} $, for the fabricated chip. For circuit layout viewpoints, one could thereby instantiate, $\alpha '_{90^{\circ } } \bigcup \beta '_{90^{\circ } } \parallel \delta '$, where $\delta '\parallel \delta $ and $\delta \rlap{$\hspace{4pt}/$}\equiv \delta '$.

\vspace{10pt}
\begin{flushleft}
\hspace{35pt}\includegraphics[width=20.9mm, viewport=0 0 10 57]{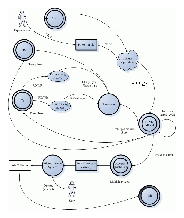}\\
\end{flushleft}
\vspace{-4pt}
\noindent{\footnotesize{\textbf{Figure 1.1.} A data flow representation of PTVD-SHAM system and I/O behaviour for storing sequences of variant data-time bits on a computer database system. Excited charge, $e$, and active regions of the $\alpha ,\; \beta , \; \delta $ poles of the chip are deemed to possess light emitting property via e.g., edge-emitting light emitting diode (ELED). Also, actors of human-computer interaction (HCI) play a vital role in collecting data during PTVD-SHAM's time-data analysis.
\smallskip

\small \noindent The complementary pole of $\delta $, is $\delta '$, whereas the latter or former are perpendicular to junctions $\alpha '_{90^{\circ } } $ and $\beta '_{90^{\circ } } $, geometrically (Fig.$\,$2.1\emph{d}, \S\ref{section2}).

Associatively, as a preamble to Fig.$\,$2.1, \S\ref{section2}, the data flow diagram on p.$\,$7 embeds into its structure, the states of storable time and data in release in form of sequences of bits. This diagram emphasizes on (1.9)'s I/O analog behaviour despite of depository vector's pattern-complexity of inputs/outputs between the chip's logic gates. A loop is generated solely for refreshing time-data cycles via time phase, $\phi _{2} $. Moreover, this loop, is for storing light purposes which is already apparent to scientists that, storing photons as minute energy packets of electromagnetic radiation into a memory cell is somewhat impossible unless, done in some quantum interference experiment.

It is crucial to not confuse `time-data' with `time-varying data' visualized by Joshi \& Rheingans in~\cite{43-Joshi}, which deals with data in motion and information concerning the kinematics of an arbitrary body in our universe. Even though these people propounded time as individual steps of event counts and snapshots taken for large data-sets (abstract of Ref.~\cite{43-Joshi}), `time' in the current chip by contrast, denotes to all events of past, current and future, `simultaneously'. `Data' however, indicates to all information related to time and space inclusively, once `time of time-date' is manifested for that particular event's information tuneable to any time location.

So by comparison, on the one hand, `data' itself in this chip is not tuneable autonomously unless dealt with the operating computer user (operator), while on the other hand, `time' is tuneable autonomously in this chip without a human user interaction (operator dependant; HCI dependant).

PTVD-SHAM elucidates a new generation of methodologies to store `time-data' parallel to, or, even into light (e.g.,~\cite{50-Howell}), once a few photons are looped between memory's cavity system and quantum states of photovoltaic effect when the cavity for excited electrons reaches a thermal level above Fermi level (see pp.$\,$9 and 11). In this case, the total energy of either form of matter representing bits of information satisfying relation (1.6), could be preserved during data transactions between all memory locations in a cyclical set of: 1- time frame, 2- data frame and 3- converted frame of electrons-light/light-electrons, across the memory channels. These memory channels are of charge material for the non-photonic material (Fig.$\,$2.1\emph{c}, \S\ref{section2}), and optical micro-filters for the path of light installed discretely as an extension to polar sites of $\alpha $, $\beta $ or $\delta $ (Figs.$\,$1.2; 2.1 of \S\ref{section2}). These chip-scale filters govern the intensity, deflections and reflections of a photonic material that carries information whilst being propagated through the system. The photonic material for optical buffering~\cite{27-Wiki}, mainly deals with the cavity system, absorbers and filters with a determination factor of light interference and deflection course during operation.

It is conceivable that a chip-scale micro-Faraday filter with apposite direction to an optical fiber plays an important role in-between poles of $\alpha $ and $\beta $ installed into zone ${\rm Z}'$ of Fig.$\,$2.1\emph{d}, \S\ref{section2}, isolated from the gate electrodes of either pole in the centre i.e. poles of $\delta $. The optical process of either micro-Faraday filter whilst having their own angle of light deflection, the filters' beam reflection is to preserve interference and coincidence of photons coming from the optical buffer to the receiving unit. The schematic representation of these filters is given in Fig.$\,$1.2, where the method generally maps from the diagrammatic setup of~\cite[xi]{27-Wiki} and~\cite{49-Knappe}, now into micro chip-scale measurements. Designing the micro-optical fiber complies with light spectra of high/low intensity applications akin to edge-emitting LEDs (ELEDs) layout, does benefit from the following classical equation~\cite{46-Senior}. The value of the normalized frequency $\nu $, hereinafter referred to as the $\nu $ number in an optical fiber path is given by
\smallskip
\begin{flushleft}
$\nu =\left(2\pi a\lambda ^{-1} \right)\sqrt{n_{1}^{2} -n_{2}^{2} } {\rm \; },{\rm \; }\sin \theta _{1} \sin \theta _{2} ^{-1} =cv^{-1} =n_{2} n_{1}^{-1} {\rm \; ,\; or,\; }\sin \theta _{1} n_{1} =\sin \theta _{2} n_{2} .$
\end{flushleft} \vspace{-16pt}
\begin{flushright}
(1.12)
\end{flushright}
\vspace{-6pt}
where \textit{a}, is the radius of the fiber core and $\lambda $, is the wavelength of the light propagating in vacuo. The $\nu $ number of the normalized frequency in (1.12), is a parameter for defining the construction of the optical fiber. In a region in which the $\nu $ number is less than 1.57, only a single mode of propagation exists. This region is referred to as: a single mode region. When the $\nu $  number increases above 1.57, a greater number of modes appear in proportion to the increase of the $\nu $ number~\cite{45-Kapany}. Base upon Snell's law, $n_{1} $ and $n_{2} $ represent the refractive index of the optical fiber medium, relevant to the principles of optical fiber communications, computing how much the speed of light, $v_{1} $ as $c\approx 300000{\rm \; }{\rm kms}^{-1} $, is reduced inside the medium in terms of $c\to v_{2} =v$. Other essential optical measurements in the experimental phase of the PTVD-SHAM project should obey the standards of~\cite{46-Senior} (see also, \S\ref{section6}).

For the chip's time-data in a photon, the design and integration methods of~\cite{49-Knappe}, is associated with the current optical buffering system resulting `Faraday effect', which is based on the classical theory of `Zeeman effect' as ensued in regard to Hyp.$\,$4.2, \S\ref{section4}.

The manifestation procedure, or, Megabit-pixel per second (${\rm Mbp}\: \rm s^{-1} $) recording system of `image logic', can be done by incorporating charge-coupled device (CCD)-chipset layer and red-green-blue (RGB) filter, installed above the optical buffer layer. The classical equation is
\vspace{-20pt}
\begin{flushleft}\[\zeta  = {\cal V}{\bf B}\ell ,\:\:\:\:\:\:\:\:\:\:\:\:\:\:\:\:\:\:\:\:\:\:\:\:\:\:\:\:\:\:\:\:\:\:\:\:\:\:\:\:\:\:\:\:\:\:\:\:\:\:\:\:\:\:\:\:\:\:
\:\:\:\:\:\:\:\:\:\:\:\:\:\:\:\:\:\:\:\:\:\:\:\:\:\:\:\:\:\:\:\:\:\:\:\:\:\:\:\:\:\:\:\:\:\:\:\:\:\:\:\:\:\:\:\:\:\:
\:\:\:\:\:\:\:\:\:\:\:\:\:\:\:\:\:\:\:\:\:\:\:\:\:\:\:\:\:\:\:\:\:\:\:\:\:\:\:\:\:\:\:\:\:\:\:\:\:\:\:\:(1.13)\]
\end{flushleft}
where, magnetic field \textbf{B} of the equation, is measured in Teslas and is excluded from the \textbf{B}-field of CCNTs, thus, this field is in enclosure to the magnetic field of the magnets surrounding the vapor cell; Symbol ${\cal V}$, is the Verdet constant for the material measured in units of radians per Tesla per meter; $\ell $, is the length of the path (in meters) where the light and magnetic field interact; $\zeta $, is the angle of rotation measured in radians.

For isolating the magnetic field in the frame of alkali vapor cell with the surrounding magnets from the optical fiber environment, we could also state
\vspace{0pt}
\begin{flushleft}
\(\emph{\textbf{A}}_{iso} =\Delta \ell b=2\pi b\left(b+\ell \right)\left|\emph{\textbf{A}}_{iso} \bigcap \mathbf{B}=0\right., \mathrm{where} \; \;  \emph{\textbf{V}}_{iso} =2\pi r_{\mathrm{C}}\mathop{{\rm {\int}}}\nolimits_{_{\min A} }^{^{\max A}}f\left(\emph{\textbf{A}}_{iso} \right)\mathrm{d}\emph{\textbf{A}}_{iso}
=2\pi \left\langle r\right\rangle \ell b,{\rm \; }b\ll r_{\mathrm{C}} ,\)

$\exists b\wedge \mathbf{B}\to 0=\left. b\right|b\in \mathop{\wedge }\limits_{i=0}^{n} b_{i} \backepsilon \left\{\begin{array}{l} {{\rm if}{\rm \; \; }\forall b\wedge \mathbf{B}=\mathbf{B}{\rm \; }, \therefore \emph{\textbf{V}}_{iso} \rlap{$\hspace{4pt}/$}\equiv \emph{\textbf{V}}_{iso} \Rightarrow \left(\mathbf{B}_{iso} :\mathbf{B}\right)\equiv \frac{\mathbf{B}_{iso} }{\mathbf{B}} =\neg \mathbf{B}_{iso} \in \mathbf{B}{\rm \; }}\\ {{\rm if}{\rm \; \; }\forall b\wedge \mathbf{B}\to 0=b{\rm \; },\therefore \emph{\textbf{V}}_{iso} \equiv \emph{\textbf{V}}_{iso} \Rightarrow \left(\mathbf{B}_{iso} :\mathbf{B}\right)\equiv \frac{\mathbf{B}_{iso} }{\mathbf{B}} =1} \end{array}\right\}.\:\:\:\:\:\:\:\:\:\:\:\:\:\:\:\:\:\:\:\:\:\:\:\:\:\:\:\:\:\:\:\:\:\:\:\:\:\:\:\:\:
\:\:\:\:\:\:\:\:\:\:\:\:\:\:\:\:\:\:\:\:\:\:\:\:\:\:\:\:\:\:\:\:\:\:\:\:\:\:\:\:\:\:\:\:\:\:\:\:\:\:\:\:(1.14)$
\end{flushleft}

\noindent $\emph{\textbf{A}}_{iso} $, is the slice of isolated surface area (change of rectangular), whereas $\emph{\textbf{V}}_{iso} $, is the isolated cylindrical volume (shell method) from \textbf{B}-field, expressing the involved magnets for the alkali vapor cell in isolation to the involved optical fiber. Symbol $b$, denotes the thickness of the shell and, $r_{\mathrm{C}} $, is the circular radius expressing the involvement of $b$, being excluded as $b\wedge \mathbf{B}\to 0=b$. Otherwise, $b$, is being included as, $b\wedge \mathbf{B}=\mathbf{B}$, which thereby deduces the volume isolation as not being isolated or, an unsuccessful attempt of \textbf{ B}-field isolation from other circuitry components, paradoxically; i.e. non-isolated\textbf{ B}-field, or, $\neg \mathbf{B}_{iso} $ which is of \textbf{B}-field type. An `AND closure operator', or, $\mathop{\wedge }\limits_{i=0}^{n} b_{i} $ is here to assist the problem for all limit conditions of $b$ not merely being existential to one $b$ (denoted by symbol $\exists $, with a commencing zero index identity). Hence satisfying all conditions of $b$ in (1.14), is when $b_{i} $ intersects with the rest of $b$-elements for all \textbf{B}-field isolated/included-ratio, $\mathbf{B}_{iso} :\mathbf{B}$. Similarly, the interference of \textbf{E}-field with $b$ should also imply to logical conditions of $b$, as given.

\bigskip
\vspace{62.5mm}
\begin{flushleft}
\includegraphics[width=16.5mm, viewport= 0 0 10 10]{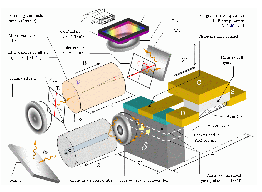}\\
\end{flushleft}

\noindent{\footnotesize{\textbf{Figure 1.2.} A schematic representation of one of the poles, here, the $\delta $-pole, incorporating micro-Faraday filter and optical paths between the active layer of the pole and a CCD. In this design, a plano-convex lens focusing the edge emitting beams of light and a collimated lens, aim to collimate light in parallel lines to the filter. For other chip's poles, if involved, optical coupling and path from source to CCD is required. ELED's light output is of incoherent versus lased-light, based on spontaneous emission versus stimulated emission.}

\smallskip
\small In continue, since $\left\langle r\right\rangle $ is the average radius of the shell, the outer radius is, $\left\langle r\right\rangle +2^{-1} b$ and thus the inner radius would be, $\left\langle r\right\rangle -2^{-1} b$. As a result, the volume of the shell is: (\textit{Volume of micro-Faraday filter}) -- (\textit{Volume of micro-vapor cell}), i.e., the deduction made on Pred.$\:(1.14)$. This theoretical concept of isolating microchip components from one another is believed to be essential, especially when the CCD and lenses are too, associated during and after the fabrication process.

\section{Description of the SHAM Chip}
\vspace{2pt}
\label{section2}
\markboth{}{Description of the SHAM Chip}

The technology incorporates in its circuital fabrication or anatomic phase,

\vspace{6pt}
\noindent --------------- \textit{Firstly}: the substrate possessing layers of sapphire insulation; subsequently, in the super-helical circuit's structure, one buffer layer out of two having GaN thus, resembling one pole of the combined switchable transistors to some HEMT heterostructure architecture, in principle. The air-core spatial aspect in measurement, explicates a free zone which does not allow the involvement of terminals expressed as gate, source or drain in the initial fundamental structure of transistors, here, containing quantum wells in the chip's structure. Therefore, one pole has a free zone ${\rm Z} $; let this be defined for the upper pole of this memory chip, and let the remaining pole (lower pole) be a Faraday effect zone of ${\rm Z} '$, for this pyramidal circuit (see also Fig.$\,$1.2, \S\ref{section1}). Note that, the buffer in either pole shall remain consistent for any incoming data such that, the carbon nanotube sustains the best location of data at any point of data loss probable occurrence (escape of data flow or due to some technical problems on voltage bias), reciprocally. A substitution for the HEMT technology in this invention-model is proposed methodologically as MOSFET in aim of engineering 2DEG, where the latter is deemed to possess high transconductance measured in Siemens suitable for common memory design and architecture incorporating CNTs (e.g.,~\cite{06-Choi}). However, MOSFETs fail to candidate themselves as high memory devices discussed in~\cite{06-Choi}, and the suggestion on an alternative solution is the current technology in the perception of HEMT and coiled CNTs embodied in this invention model. The prime agent to this solution is to use a catalyst, electric conductor and ${\rm SiN _{x}}$, in regard to the insulating layer within the fabrication process. ---------------
\vspace{6pt}

\noindent --------------- \textit{Secondly}: the installation grade of a pair of carbon nanotubes in some geometry of say, e.g., double-helix with resemblance to a helical antenna used in `plasma physics', mappably. The links between their gaps (or, bonds) shall be given in terms of electron jump in its displacement between the nanotubes' membranes measured in deca of angstroms or, 10Å = 1nm multiplied by a scalar real number, $\mathfrak{r}\mid\mathfrak{r}\in\mathbb{R}$, for the occupying vector space. This measurement defines sizes of membrane strands' nodal integration in a tensor manner for the electron jump at any point plausible, probabilistically. The result of this, allows an ultimate storage of data no matter the mutation course of some particle, in this case, an electron on any level of the permittivity state for this super memory. Furthermore, the notion of indices $\left(n,m\right)$ as `chiral vector', is not acknowledgeable as, $n=m$, representing the helix nanotube material as `purely metallic' for CCNT perceived by Lambin \textit{et al.}\textit{ }of~\cite{21-Lambin}. Thus, $n\ne m$ is the most adequate in preserving a semi-conducting electrical property in practice (generalized by e.g., Lambin \textit{et al.} in~\cite{21-Lambin}). The junctions between CCNTs from one staircase step to another per catalyst, is of $\left(n,m\right);m-n=3q$, where \textit{$m$}, $n$ and \textit{$q$ }are integers, showing the transition, metallic$\to $semiconducting$\to $metallic~\cite{39-Umeno}, forming in this case, a CNT diodic junction (see Fig.$\,$2.1\emph{a}, and~\cite{08-Cohen}) done by a linear junction between armchair and zigzag. The CCNT diode junctions, ideally represent photovoltaic effect when the cavity for excited electrons reaches a thermal level above Fermi level measured in Electronvolts (eV). This effect occurs, once the excitation of electrons and their overall conduction process are in place, hence converting light into electrical energy. The conversion of light occurs, once the accumulation of energy valance begins to perform quasi-Fermi levels (or, the sum of quasi-quantized energy states) for the majority and minority of carriers, mobile within the cavity's environment. ---------------

\vspace{6pt}
In contrast to CCNTs fabrication, the growth direction of pillar-shaped CNTs (observe Fig.$\,$2.1) follows the direction of the electric field in the plasma enhanced chemical vapor disposition (CVD) process discussed by Ren \textit{et al}. in~\cite{32-Ren} and exemplified by Umeno \textit{et al.}\textit{ }in~\cite{39-Umeno}. In this case, even under high axial tension, CNTs should satisfy a vertical metallic interconnection course of integration whilst preserving a chiral vector value of $n=m$ or $\left(n,m\right);n-m=3q\to 0$ as being an armchair type remaining metallic or, $\left(n,n\right)$ discussed by Collins \& Avouris in~\cite{09-Collines}.

\vspace{6pt}
In contrast to CCNTs fabrication, the growth direction of pillar-shaped CNTs (observe Fig.$\,$2.1) follows the direction of the electric field in the plasma enhanced chemical vapor disposition (CVD) process, discussed by Ren \textit{et al}. in~\cite{32-Ren} and exemplified by Umeno \textit{et al.}\textit{ }in~\cite{39-Umeno}. In this case, even under high axial tension, CNTs should satisfy a vertical metallic interconnection course of integration whilst preserving a chiral vector value of $n=m$ or $\left(n,m\right);n-m=3q\to 0$, as being an armchair type remaining metallic or, $\left(n,n\right)$ discussed by Collins \& Avouris in~\cite{09-Collines}.

In continue, the threshold voltage $V_{th} $, increases versus one of the gate voltages when $I_{d} $ remains constant, hence holes from the CNT are injected to the oxide-nitride-oxide (ONO) thin film between CNT and the gate electrode. It is deemed that from the electric field between each CNT and the gate electrode, the induced charge density $\sigma $,  increases with proximity to polar CNTs of the chip (refer to the lower right graph of Fig.$\,$2.1\emph{b}), discussed by Choi \textit{et al.} of~\cite{06-Choi}. Moreover, localized charge distribution enables charges to be induced into the nitride film of the ONO thin film due to the high electric field distribution from localized CNTs  (Fig.$\,$2.1\emph{c}), and charges trapped in localized areas of the ONO thin film may be dispersed concerning drain current $I_{d} $, for either pole. This behaviour is normalized by utilizing metallic CNT adjacent to CCNTs, splitting the induced charge into different traps of the ONO as a choice of path for charge $e$ (the continuation to this, is subject to \S\ref{section4}). Hence, a CNT with a diameter of 3 nm for an ONO thin film between the CNT and the gate electrode, being regarded as one single layer with an effective dielectric constant of 3, the calculated electric field for an applied gate voltage of 5 V would be, 970 V ($\rm \mu m) ^{-1}$~\cite{06-Choi}.

\vspace{1pt}
\begin{flushleft}
\includegraphics[width=23.4mm, viewport= 0 0 10 50]{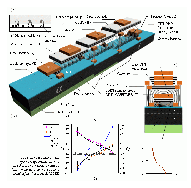}\\
\vspace{-8pt}\includegraphics[width=18mm, viewport= 0 0 10 52]{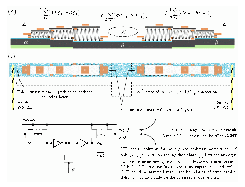}\\
\end{flushleft}
\vspace{-3pt}

\noindent{\footnotesize\textbf{Figure 2.1 }(\emph{a}) A three-dimensional perspective view; (\emph{b}) cross section view including a 4-dimensional chart of charge interactions between poles and 2DEG layering, with; (d) a complete pyramidal view of the involved poles of the general design; (\emph{e}) top view of (\emph{d})  plus a diagrammatic solution over the chip. The displayed components in representation are sectional to the design created by the author whilst; the layering and dimensions of electrodes could obey models of Choi \textit{et al.}~\cite{06-Choi}; Krämer \textit{et al.}~\cite{19-Kramer} and Okita \textit{et al.}~\cite{26-Okita}, in description; (\textit{c}) represents the memory layer nanodots to store data discussed by Choi \textit{et al.}~\cite{06-Choi}.
\vspace{8pt}

\small
In other words, the exemplified gate voltage forms a strong \textbf{E}-field which is enough to induce a Fowler-Nordheim tunneling in the memory's CNFET application set using MOSFET technology. Furthermore, collapse of current or `gate lag' problem is considered to be reduced between the division points. This reduction occurrence, is due to the proportional role of the CNT interconnection against CCNT in normalizing the electric current between the HEMTs from either pole, connecting source and drain\textit{ }electrodes to the gate\textit{ }electrode via contacting memory cell layer, simultaneously.

Before the depletion stage of current $I_{d} $, in its time response measurement for either pole as, $\alpha \left(I_{d} \right)$ and $\beta \left(I_{d} \right)$, the carbon Fermi energy in diodic points (contacts) occurs wherever CNTs join. These CNTs join each other with different electrical property (see nanotube joints and compare this with~\cite{08-Cohen}), exhibiting quantized charges with sizes of $\le 10$ nm at room temperature. This type of quantization attracts structures for dynamic storage devices\textit{ }and\textit{ }data receivers in conventional computers and thus, resonators here as, e.g., super-RAMs and telecommunicators. The reason is the model's practical role for a real high to low p.d.-drag, defining the `\textit{waterfall effect}' in the chip's application.

The diode-contacts mentioned formerly, denote the `voltage drop versus increase at the ends of the circuit in compensation' from both sides i.e., the cyclical state of electric current from either pole as, $\alpha \left(I_{d}\right)$ or $\beta \left(I_{d} \right)$, correspondingly. The prevention factor on electric current coming from either CNFET with their base-CNT, is to prevent leakage problems thus acting like valves in-between $\alpha $ and $\beta $ of the chip. Nevertheless, once the overflow occurs, the voltage drop on one side, does not mean the loss of data since on the other side, as the other CNFET pole, provides the same data from drain electrode reversibly. This is due to the structural representation and orientation degree of the CCNTs and catalysts.

The following is in conjunction with Fig.$\:2.1$\emph{a} materials and components as pointed out by number, i.e., a mere theoretical embodiment of the invention-model, where this integrated conducting/semiconducting device set is comprised of:

\begin{enumerate}

\item [{1-}] Copper (Cu) gate as a conductor from HEMT-to-helix or core and vice versa (the second as `helix' or coil, is done by priority for the privileged places to store electrons, otherwise, it is the prioritization of the conduction grade of electrons).

\item [{2-}] Unintentionally doped (UID) AlGaN. \\ The thickness size varies from, \hspace{-3pt} $\theta \in \left({\rm 6.67\; nm\; ,\; 25\; nm}\right]$, in the layer's geometric space measurement for its staircase thickness, based on ratio $\theta _{2} :\frac{\theta _{1} }{2} ,\frac{\theta _{1} }{3} :\theta _{3} $ wherefrom left as middle to right as minimum-$\theta $, represents ratios, 2nd:1st and, 1st:3rd descending stair, respectively.

\item [{3-}] ${\rm SiN _{x}}$ insulating layer within the staircase core of the PTVD-SHAM Chip.

\item [{4-}] Gate: Titanium (Ti) or Gold (Au). This applies to thick gates' principles for width and length measurements.

\item  [{5-}] Memory cell e.g., $\left({\rm SiO}_{2} /{\rm Si}_{3} {\rm N}_{4} /{\rm SiO}_{2} \right)$ with a thickness of \scriptsize {$ \sim $}\small 28-30 nm.

\item  [{6-}] The symmetry of MOSFET/HEMTs as `upper pole' (frontal portion is the memory cell).

\item [{7-}] Unintentionally doped (UID) GaN buffer layer. Size: 1$\mu$m for its thickness.

\item [{8-}] Sapphire substrate symmetric to MOSFET substrate, is for stable current against alternate current (AC) between terminals and the incorporated CNTs of this chip which possess high electrical conductivity (recall pp.$\,$ 11 and 12). The thickness of this layer varies up to 350 $\mu$m.

\item [{9-}] Cavity layer inclusive to the paired CNTs between the base terminals.

\item [{10-}] A novel mechanism representing electro-migration and multilevel metallization, here for vertical metallic interconnection (purely metallic CNT pillar construct with a chiral vector value of $n=m$, e.g., pillar-CNT, part \textit{b}). This mechanism basically appreciates the model and utilization layout claimed by Greenberg \textit{et al.} in~\cite{15-Greenberg}. Methods for bundling the pillar-CNTs attached to the staircase electrodes using electrophoresis, is claimed and discussed by Kim \textit{et al.}~\cite{16-Kim}. Despite of number of turns, a CCNT in association with pillar-CNTs has tube and coil's minimum diameter of \scriptsize {$ \sim $}\small30 nm and 300 nm, respectively, with a theoretical inductance measured in micro-Henries.

\item [{11-}] A 4-dimensional chart of space-time charge interactions for all poles; spatial axis $\lambda _{1} $ is measured in microns and, $\lambda _{0} ,\lambda _{2} $, as maximum-to-minimum geometric length units of (1.11), where Hertzian concentration of 2DEG by time function $f\left(t\right)$, is measured in Hz. These considerations are satisfied by logarithmic interactions between poles, where the spatial occupation multiplied by time-fractal $t^{-1} $, generates the speed of interaction-mean for the circuit's total charge. Delta pole, $\delta $, is $f\left(t\right)$-dependant evaluated by time phase $\phi _{2} $, compared to pole's lambda spatial property.

\item [{12-}] Zone ${\rm Z'}$, serving the optical buffer and relevant micro-filters explained earlier on pp.\,9 and 10, \S\ref{section1}; debates on the conditional planar or architectural extensions to the chip's poles.

\end{enumerate}

\section{Functions and operations }
\label{section3}
\vspace{4pt}
\markboth{}{Functions and operations}

The time-varying \& data super helical access memory (PTVD-SHAM)'s main functions, satisfying a state of conservation of all classical Boolean logic \& image-logic states, in a non-volatile versus volatile mode conversely, is categorized and conceptually assigned by
\vspace{6pt}
\begin{enumerate}
\item  Incorporating HEMT and MOSFET Logic for memory layers.

\item  Incorporating, waterfall wave filters, here as time-purposed crystal oscillators; a combination of CCNT and Capacitors as series band-pass filters to form a passageway of normal frequencies between I/O, thus attenuating frequencies outside the range during operation with an advantage of freedom of choice of precise time and location for data. This type of oscillation would be more advanced than say, e.g., crystal momentum conserved in interactions between electrons and phonons in a crystal~\cite{18-Kittel}.

\item  Integrating catalysts for best conductivity, reducing complexity on CCNT configuration as synthesized between catalysts at specific staircase sites. This is for regularizing thee black body effect i.e. an object that deflects and thus controls energy accumulation by distributing it into other layers of adjacent waterfall chambers. The energy distribution prevents `thermal radiation overlapping', including the conversion of photonic states of data into charge states of data.

\item  Satisfying cavity's quantum property via pillar-CNTs, memory cells and coiled CNTs as CNFETs and CCNT-FETs, respectively.
\end{enumerate}
\vspace{6pt}
The operation module that involves the electro-migration throughout the functions mentioned above, for the circuit's electronic components would then associate to operational conductivity problems as follows:

\vspace{6pt}
\noindent --------------- Once the delivery of electrons to the drain electron is made from the polar layers and the base carbon nanotube, it is then considered the mobility of electrons between the helical carbon nanotubes averaging between the highest mobile electrons (ballistic transport of hot electrons in the 2DEG layer according to relation 1.2, \S\ref{section1}), as the ones attaining the soonest. In return, the typical electrons being conducted via base-CNT therefore, are considered the soonest otherwise the last ones in arrival. Hence, the delayed charges of electrons (holes) are stored to the memory cells' last addresses or the farthest memory location to base-CNT (or, 1st rising stair, where biggest helical CNT resides). Consequently, the soonest ones in terms of first-in-first-out (FIFO) are the ones already stored in the memory cell's address e.g., leftmost part of the upper pole's wafer construction against the notion of the first popped-out data in the rightmost part of the upper pole's wafer. This is done when writing data operation is enabled via PTL logic setup.  ---------------

\vspace{6pt}
This type of conduction, prevents latency and other problems in accessing data by averaging the time of data-dispatch, store and access, now prioritized as: `Where is best to deposit data pre-per second'?

This so-called `pre-time attribute' is forwarded for a pre-optimized prior location and time for data being stored and thus accessed functionally/instantaneously via helical CNTs. The time-data accessability is a resolved issue due to the existence of the helical structure of the incorporated-paired CCNTs. The mutual conductance or transconductance as described in the previous paragraphs on late and early carriers here, electrons, from both HEMT poles to the centre (or, the staircase where helical CNTs embed in), is calculated by
\vspace{-10pt}
\begin{flushleft}
$$\left\langle g_{m} \right\rangle _{\alpha ,\beta } =\frac{1}{j} \sum \frac{\Delta I}{\Delta V} \; \; \textrm{for} \; \; \alpha \wedge \frac{1}{j} \sum \frac{\Delta I}{\Delta V}, \; \textrm{for} \; \beta , \; \textrm{such that}, \:\:\:\:\:\:\:\:\:\:\:\:\:\:\:\:\:\:\:\:\:\:\:\:\:\:\:\:\:\:\:\:\:\:\:\:\:\:\:\:\:\:\:\:\:\:\:\:\:\:\:\:\:\:\:\:(3.1)$$

\vspace{-6pt}
$$\left\langle g_{m} \right\rangle _{\alpha ,\beta } \equiv \frac{1}{j} \sum \limits _{i=0}^{j\ge 2}\left(\frac{\Delta I}{\Delta V} \right)_{i}  =\frac{\Delta I_{ds,\mathrm{CNT}} }{\Delta V_{gs,\mathrm{CNT}\left(\alpha ,\beta \right)} } =\frac{\Delta I_{ds} +\left(\Delta I_{\mathrm{CNT}_1} -\Delta I_{\mathrm{CNT}_2} \right)}{3\Delta V_{gs} +\Delta V_{\mathrm{CNT}\beta } +\Delta V_{\mathrm{CNT}\alpha } } \; \; \; \; \; \; \; \; \; \; \; \; \; \; \; \; \; \; \; \;  \; \; \; \; \; \; \; \; \; \; \; \; \; \; \; \; \; \; \; \; \; \; \; \; \; \; \; \; \; \; \; \; \; \; \; \; \; \; \; \; $$ \end{flushleft}\vspace{-16pt}
\begin{flushright}$$=\frac{g_{m1} +\Delta g_{m\mathrm{CNT}} }{2} , \; \; \; \; \; \; \; \; \; \; \; \; \; \; \; \; \; \; \; \;  \; \; \; \; \; \; \; \; \; \; \; \; \; \; \; \; \; \; \; \; \; \; \; \; \; \; \; \; \; \; \; \; \; \; \; \; \; \; \; \; \; \; \; \; \; \; \; \; \; \; \; \; \; \; \; \; \; \; \; \; \; \; \; \; \; \; \; \; \; \;  \; \; \; \; \; \;(3.2) $$\end{flushright}
\vspace{-2pt}
where the latter and former relations initially come from the transconductance formula in field effect transistor's general cases, parametrically,
\vspace{-14pt}
\begin{flushleft}
$$g_{m} =\left. \frac{\Delta I_{ds} }{\Delta V_{gs} } \right|_{{\rm \; }V_{ds} =\mathrm{constant}}.\; \; \; \; \; \; \; \; \; \; \; \; \; \; \; \; \; \; \; \;  \; \; \; \; \; \; \; \; \; \; \; \; \; \; \; \; \; \; \; \; \; \; \; \; \; \; \; \; \; \; \; \; \; \; \; \; \; \; \; \; \; \; \; \; \; \; \; \; \; \; \; \; \; \; \; \; \; \; \; \; \; \; \; \; \;  \; (3.3)$$
\end{flushleft}
\vspace{2pt}
As if said that, the distribution in (3.2) for the electric current is expressed by $\left\langle g_{m} \right\rangle $, which is the distributed ratio of all current, $I$, on the output port between the `helical CNTs and electrodes' as, $\Delta I_{ds,CNT} $ and, $\Delta V_{gs,\mathrm{CNT}\left(\alpha ,\beta \right)} $, denoting that all voltages on the input ports of the source and gate electrodes, are followed by base-CNT voltage, $V_{\mathrm{CNT}\beta } $ and $V_{\mathrm{CNT}\alpha } $, in terms of, $\Delta V_{gs,\mathrm{CNT}\beta } $ and $\Delta V_{gs,\mathrm{CNT}\alpha } $, respectively. Hence, for both poles in respect to their carriers' current direction with respect to their p.d. occurrence, therefore, relation (3.2) becomes the most privileged in practice.

\section{The role of Gaussian curvature and the concept of \\ electrons in B-field normal to external E-field via \\ Maxwell-Lorentz theory in Minkowski's space and \\ time universe}
\markboth{}{The role of Gaussian curvature}
\label{section4}

In cases where in the field of induction, a study of an electron in motion with alignment to the S and N poles of the CCNT is being prioritized, the study of other bodies as mobile electrons coming from quantum confinement in HEMT cases, are too reciprocally prioritized in product terms. The variations of energy states with the external fields are reflected in the electrical and thermal conductance of the CNT with chiral vector unit $n=m$. The CNT as `pillar CNT', is vertically constructed between the electrodes of either pole. The number, the heights, and the positions of the conductance peaks, are strongly dependent on the external fields where \textbf{B}-field and \textbf{E}-field are visited. This condition is satisfied by the following electromagnetic process initially elicited from the theoretical notion of Minkowski's space and time,
\vspace{2pt}
$${\rm {\mathfrak E}}\equiv \left(e\sqrt{-1e_{2} e'_{2} } \right)\dot{t}, \; -ee_{1} \dot{t}=\frac{\bar{\Sigma }\cdot \Sigma }{-ee_{2} \dot{t}}, \; \bar{\Sigma }\equiv \frac{\Sigma _{x} +\Sigma _{y} +\Sigma _{z} }{3} =\frac{1_{x} +1_{y} +1_{z} }{3} =1_{\tau }, \; \Sigma =1 ,$$ \begin{flushright}(4.1)\end{flushright}
\vspace{-4pt}

\noindent where $\dot{t}$, is the time factor equal to $\frac{\mathrm{d}t}{\mathrm{d}\tau } $, or simply, $\dot{t}^{-1} =\sqrt{1-v^{2} c^{-2} } $, defining `relativistic time intervals' between the storing events of data. Ergo,
\vspace{2pt}
$$\therefore \bar{\Sigma }\cdot \Sigma =1_{\tau } \cdot \Sigma =e^{2} e_{1} e_{2} \dot{t}^{2} , \; \textrm{whereby}, \;
\therefore \bar{\Sigma }\cdot \Sigma =1_{\tau } \cdot \Sigma =\frac{e^{2} e_{1} e_{2} t^{2} }{1-v^{2} c^{-2} } ,\; \; \; \; \; \; \; \; \; \; \; \; \; \; \; \; \; \; \; \; \; \; \; \; \; \; \; \; \; \; \; \; \; \; \; \; \; \; \; \; \; \; \; \; \; \; \; \;$$ $$\because t=\frac{\tau }{\sqrt{1-v^{2} c^{-2} } } \;  \mathrm{and} \;  1_{\tau } \stackrel{\frac{x_{1} }{ct_{1} } \wedge \frac{x_{2} }{ct_{2} } }{\longrightarrow}\frac{\tau }{t_{1} } ,\frac{\tau }{t_{2} } \;  \mathrm{iff} \;  \dot{t}^{2} \ne \dot{t}^{2}, \; \; \; \; \; \; \; \; \; \; \; \; \; \; \; \; \; \; \; \; \; \; \; \; \; \; \; \; \; \; \; \; \; \; \; \; \; \; \; \; \; \; \; \; \; \; (4.2)$$
\vspace{6pt}

\noindent \noindent wherein this case $\dot{t}^{2} \ne \dot{t}^{2} $, obeys Russell's Paradox (e.g.,~\cite{31-Quine};~\cite{34-Russell}) as a `free variable' satisfying global time scenarios on the current frame of time ratios, due to the probabilistic existence of electron of the charge, $e_{1} $ for $e$ of product $e^{2} e_{1} $ and, $e_{2} $ for the same $e$ of product $e^{2} e_{2} $, in the context of simultaneity. Time, $t$ of $c$ in $\frac{x_{1} }{ct_{1} } \wedge \frac{x_{2} }{ct_{2} } $ as stationary time in (4.2), is due to the definition of speed of light, $c$, an invariant speed $\approx 300000 \: \textrm{kms}^{-1} $.

In other words, time entanglement of past, current and future events for data product value in terms of data constituent, hereby as, one entangled $e$ in a probabilistic flow of entangled and non-entangled electrons (compare with quantum time entanglement of electrons~\cite{24-McGuire}), could be expressed as,

\vspace{-8pt}
$$\frac{\mathrm{d}t}{\mathrm{d}\tau } \ne \frac{\mathrm{d}t\leftrightarrow \tau }{\mathrm{d}\tau \leftrightarrow t} {\rm \; }\textrm{for}{\rm \; }e\in e^{2} e_{1} \wedge e^{2} e_{2} \Rightarrow \therefore \dot{t}^{2} \ne \dot{t}^{2} {\rm \; }\textrm{for}{\rm \; }e\in e^{2} e_{1} \wedge e^{2} e_{2} .\:\:\:\:\:\:\:\:\:\:\:\:\:\:\:\:\:\:\:\:\:\:\:\:\:\:\:\:\:\:\:\:\:\:\:\:\:\:\:\:\:\:\:\:\:\:\:\:\:\:\:\:\:\:\:\:\:(4.3)$$

\vspace{3pt}
Let time $t$ in the `time dilation' concept, represent relativistic time, and $\tau $ be the time at rest or `proper time' being validly defined in Einstein's theory of SR. Thus, specifying this proper time via implication relation elicited from $1_{\tau } $ in (4.2) for bodies $e_{1} $ and $e_{2} $, respectively, dilates
\vspace{-12pt}
\begin{flushleft}
$$\therefore 1_{\tau } \cdot \Sigma =\frac{e^{2} e_{1} e_{2} \left(t_{1} t_{2} \right)}{1-v^{2} c^{-2} }   ,                                                                                                            \:\:\:\:\:\:\:\:\:\:\:\:\:\:\:\:\:\:\:\:\:\:\:\:\:\:\:\:\:\:\:\:\:\:\:\:\:\:\:\:\:\:\:\:\:\:\:\:\:\:\:\:\:\:\:\:\:\:
\:\:\:\:\:\:\:\:\:\:\:\:\:\:\:\:\:\:\:\:\:\:\:\:\:\:\:\:\:\:\:\:\:\:\:\:\:\:\:\:\:\:\:\:\:\:\:\:\:\:\:\:\:\:\:\:\:\:\:\:\:(4.4)$$
\vspace{-6pt}
$$\therefore \left(\frac{\tau }{t_{1} } ,\frac{\tau }{t_{2} } \right)\Sigma =\frac{e^{2} e_{1} e_{2} \left(t_{1} \vee t_{2} \right)}{1-v^{2} c^{-2} } \; , \; \dot{t}^{2} \ne \dot{t}^{2}\; ,                                                                              \:\:\:\:\:\:\:\:\:\:\:\:\:\:\:\:\:\:\:\:\:\:\:\:\:\:\:\:\:\:\:\:\:\:\:\:\:\:\:\:\:\:\:\:\:\:\:\:\:\:\:\:\:\:\:\:\:\:
\:\:\:\:\:\:\:\:\:\:\:\:\:\:\:\:\:\:\:\:\:\:\:\:\:\:(4.5)$$
\vspace{-6pt}
$$\therefore 1_{\frac{\mathrm{d}\tau }{\mathrm{d}\tau } \tau } \cdot \Sigma =e^{2} e_{1} e_{2} \left(t_{1} \vee t_{2} \right) \; \mathrm{, \; or, \; equivalently,} \:\:\:\:\:\:\:\:\:\:\:\:\:\:\:\:\:\:\:\:\:\:\:\:\:\:\:\:\:\:\:\:\:\:\:\:\:\:\:\:\:\:\:\:\:\:\:\:\:\:\:\:\:\:\:\:\:\:
\:\:\:\:\:\:\:\:\:\:\:\:\:\:\:\:\:\:\:\:\:\:\:\:\:\:(\rm 4.6a)$$

$1_{1\to \left(x,y,z\right)t'} \cdot \Sigma =e^{2} e_{1} e_{2} t_{1} $,  $e^{2} e_{1} e_{2} t_{2} $ .            $\:\:\:\:\:\:\:\:\:\:\:\:\:\:\:\:\:\:\:\:\:\:\:\:\:\:\:\:\:\:\:\:\:\:\:\:\:\:\:\:\:\:\:\:\:\:\:\:\:\:\:\:\:\:\:\:\:\:
\:\:\:\:\:\:\:\:\:\:\:\:\:\:\:\:\:\:\:\:\:\:\:\:\:\:(\rm 4.6b)$
\end{flushleft}
\vspace{2pt}
\noindent The previous relation generates a `Newtonian attraction' in accordance with Einstein \textit{et al.}~\cite{14-Einstein et al.} on the basis of two points of mass, $m$ and $m_{1} $, describing their world-lines satisfying cases relating to $mm_{1} $ in the context of Lorentz's rest mass and relativistic mass of special relativity, correspondingly.

Let ${\rm {\mathfrak E}}$ represent an entanglement vector for electron charge $e$ and $e_{2} $ relativistically. Body charge, $e_{2} $, is deemed to be in confinement thus the notion of complex plane's engagement in theory stands relevant upon $e_{2} $.

In other words, the involvement of imaginary unit, ${\rm \imath \equiv \jmath}=\sqrt{-1} $ in these previous relations explains the presence of at least one complex plane pertinent to the emanating field sourced from CCNTs.

Once the problem is discussed in terms of the point of confinement on $e_{2} $ to the actual point(s) of probable entanglement locus expressing $e_{2} $ and its symmetry $e'_{2} $, as an entangled state of the same particle's occupying space, the product of vector ${\rm {\mathfrak E}}$ thereby suits against proper time $\tau $. Consequently, paradox  $\dot{t}^{2} \ne \dot{t}^{2} $, will emerge from this vector-form entanglement.

\vspace{-6pt}
\bigskip
\begin{flushleft}
\includegraphics[width=14mm, viewport= 0 0 10 30]{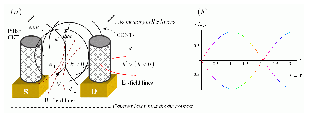}\\
\end{flushleft}
\vspace{-6pt}

\noindent{\footnotesize\textbf{Figure 4.1. }$\left(\emph{a}\right)$ The theoretical state of Gaussian curvature, $k<0$, a hyperbolic geometry from the diametric viewpoint of the CCNT's structure, from centre to circumference of the nanotube against $k'>\left(k<0\right)$ inclusive to the charge and fields' involvement; $\left(\emph{b}\right)$ curvature function, $f\left(k_{\sigma \leftrightarrow \rho } \right)$, with an evolutionary quantum charge density, $\sigma \leftrightarrow \rho $-index. See also, current Section's items: \textcolor{blue}{$\dagger$}  and  \textcolor{blue}{$\ddagger$} on the subsequent pages.}

\small
\begin{enumerate}
\item  [\textcolor{blue}{$\dagger$}] \noindent --------------- \textit{Operation}: the curves of Fig.$\,$4.1\emph{a} satisfy \textbf{B}-field and \textbf{E}-field loci where're therefore visited on charge $e$, in its electro-migration of $e^{2} e_{1} e_{2} $ on world-line at P and its symmetry $\mathrm{P}'$, for intersected charge $e$ between curvatures $k'>\left(k<0\right)$ and $k<0$. This occurs when from the drain electrode, 2D hot electron migrates into the cavity region between first point electron and second point electron of charge $e$, as $e_{1} $ and $e_{2} $, respectively. When the external hyperbola of the migration line reduces to internal hyperbola with centre M of the \textbf{B}-field lines, therefore, touching the world-line in P, achieves $\rho =\infty $ for magnitude $c^{2} \rho ^{-1} $, relevant to the debate made by Einstein \textit{et al.}~\cite{14-Einstein et al.} regarding space and time. So, the fields' interactions, permits their lines to form external curvatures in a magnitude of a displacing field which permits electro-migration on both sides of the curves at the magnetic polar points of the CCNT, such that, a selective approach (random) of the charge being stored in the memory cell becomes apparent. This selective approach for the involved charge is in manifestation, once the paradox, $\dot{t}^{2} \ne \dot{t}^{2} $, in subsequent relations sustains. This is due to the earlier vindication made upon the fields interaction course for all superimposing charges of $e$. The \textbf{E}-field lines radial to the cavity's core perform with a deflective course of $eee_{1} $ and $eee_{2} $ depending on \textbf{B}-field lines' polarity. The area of selective displacement of the charge is of elliptical volume otherwise spherical volume of space (see e.g., relation 4.18) demarcated between hyperbolic points of the fields. The eventual resultant on this particle(s) demarcation, produces `image-logic patterns' of the stored charge in terms of Moiré pattern, submitted in relation (4.21). The selective superposition state could be amplified by the catalysts between CCNTs. Fig.$\,$4.1\emph{b} curvature function, $f\left(k_{\sigma \leftrightarrow \rho } \right)$, with an evolutionary quantum charge density, $\sigma \leftrightarrow \rho $-index, represents continuous charge density of total charge, $Q$, of Gauss's law over two-to-three dimensional worldline whose curvature expands and contracts spatial dimensions between \textbf{B} and \textbf{E}-field lines. This expansion/contraction of space dimensions is done via time line as radial time distance $r$, relevant to equation (4.18), introducing contour integral results, (4.10 and 4.11), as a follow-up on Fig.$\,$4.2. The latter integration output, is adjusted for energy spectra satisfying the contraction and expansion grades of energy bands for storable and passing charges (or, transitory charges) between electrodes, cavity and memory cell layers. Colour representation of the spectra performs a relativistic Doppler shift effect (e.g.,~\cite{35-Saathoff}) from the body during curvature's geometric transformation. The charge's mass condition by its energy state relevant to the evolution mentioned earlier (Fig.$\,$4.1\emph{b}), is given in Tab.$\,$4.1. ---------------

\item  [\textcolor{blue}{$\ddagger$}] \noindent --------------- \textit{ Operation's deduction}: it is by operation deduced that, the Gaussian curvature of $k'>\left(k<0\right)$ against $k<0$ from Fig.$\,$4.1, for any problem involving all geometric conditions upon bodies in the \textbf{B}-field and active layers of the semiconductor (here as the 2DEG layer), remains intact to the notion of 1-bit logic against 2-bit logic. The latter bit form, equivalently in quantity, represents one qubit to the course of entanglement principle. This course of entanglement for one qubit, assumably occurs within the magnetic field prior to the particle's confinement. ---------------
\end{enumerate}

\small At issue, the fundamental notion to confinement is of signifying $e^{2} e_{1} e_{2}\; $, which is always equal to storage vector, $\Sigma =1$, cross product to its spatial $\Sigma _{x} $, $\Sigma _{y} $ or $\Sigma _{z} $ as a net storage product $\bar{\Sigma }$, such that 1, here, represents 1-bit in favour of some spatial dimension in form of $\Sigma _{x} $, $\Sigma _{y} $ or $\Sigma _{z} $. The latter storage product is being fulfilled, once the net direction of motion of $e$, $e_{1} $ and $e_{2} $ is clarified, i.e., when the ratio to the defined spatial directions is too evaluated, accordingly.
This $\bar{\Sigma }$ is neither of velocity vector $\vec{v}$, nor must be confused with acceleration vector $\vec{a}$ ascribed in Minkowski's space-time viewpoint, whereas, supposing that, \textbf{}arbitrarily one of the axes $x,y$ or $z$ e.g., by laws of delimitation on $z$ approaching $z=0$, eludes in form of
\vspace{-18pt}
\begin{flushleft}
$$\mathop{\lim }\limits_{\Delta z\to 0} 1_{\frac{\mathrm{d}\tau }{\mathrm{d}\tau } \tau } \cdot \Sigma =1_{x} \cdot 1_{y} =\log _{\Sigma } 1_{xy} =2\:\:\:\:\:\:\:\:\:\:\:\:\:\:\:\:\:\:\:\:\:\:\:\:\:\:\:\:\:\:\:\:\:\:\:\:\:\:\:\:\:\:\:\:\:\:\:\:\:\:\:\:\:\:\:\:\:\:
\:\:\:\:\:\:\:\:\:\:\:\:\:\:\:\:\:\:\:\:\:\:\:\:\:\:(4.7)$$
\end{flushleft}

\noindent , forming a two dimensional cell, shall store data in favour of storage vector $\Sigma $ for the exemplified dimensions $x$ and $y$ in two different stationary time scenarios, $t_{1} $ and $t_{2} $ of $\mathrm{d}t\leftrightarrow \tau $, as follows
\vspace{6pt}
\begin{flushleft}
$\therefore \Sigma ^{2} =\left\{1_{xy} =e\sqrt{e_{1} e_{2} e'_{2} t_{1} } \left(e\sqrt{e_{1} e_{2} e'_{2} t_{2} } \right)\right. $, $\:\:\:\:\:\:\:\:\:\:\:\:\:\:\:\:\:\:\:\:\:\:\:\:\:\:\:\:\:\:\:\:\:\:\:\:\:\:\:\:\:\:\:\:\:\:\:\:\:\:\:\:\:\:\:\:\:\:
\:\:\:\:\:\:\:\:\:\:\:\:\:\:\:\:\:\:\:\:\:\:\:\:\:\:(4.8)$

$\therefore \Sigma =\left\{1_{x} =e\sqrt{e_{1} e_{2} e'_{2} t_{1} } ,1_{y} =e\sqrt{e_{1} e_{2} e'_{2} t_{2} } {\rm \; },{\rm \; }1_{y} =e\sqrt{e_{1} e_{2} e'_{2} t_{1} } ,1_{x} =e\sqrt{e_{1} e_{2} e'_{2} t_{2} } \right. $ .
\end{flushleft}
\begin{flushright}\vspace{-6pt}(4.9)\end{flushright}

\noindent This eccentrically satisfies one of the spatial dimensions properly thus, articulating that $e_{1} $, could also be entangled with $e_{2} $, whilst having in position, $e'_{2} $, once the product is emerged from $1_{\tau } \cdot \Sigma $. Therefore, $e_{1} $, becomes well-vindicated in its spatial data depository characteristics for proper time interval, $\mathrm{d}t\leftrightarrow \tau $, against the occurred time factor $\dot{t}$, and $1_{xy} $, representing the normal to `velocity vector' in the hyperbolic world-line which has been already occurred at the moment of depository's bitwise transformation.

\begin{flushleft}
\includegraphics[width=14mm, viewport= 0 0 10 50]{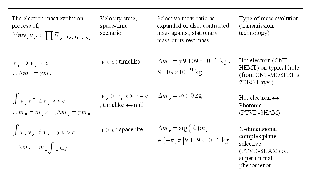}\\
\end{flushleft}

\noindent{\footnotesize{\textbf{Table 4.1. }Particle dynamics/kinematics evolution table. Symbol, $\tilde{\infty }$, merely represents complex infinity argument by Krantz~\cite{20-Krantz}  depending on the body's mass state of selecting trap sites via radius of time $r=t$, exclusively (see Fig.$\,$4.2). Symbol, $\gamma $,  represents relativistic gamma from the theory of SR, whereas $e'$, is the theoretical counterpart of charge $e$ satisfying far and close radial space and time distances between cavity and memory cell traps.}
\vspace{4pt}

\small Thus, such vectors normal to velocity vector are ignored since the occurrence of time factor $\dot{t}$ from time function $f\left(t\right)$ for $\Sigma $, is prematurely simplified into $\tau _{\Sigma } $ with area coordinates' limit to (4.11)'s time-symmetries in say, some `image logic'.

However, $\bar{\Sigma }$ could be understood with the analogy of Newton's second law, $\vec{F}=m\vec{a}$, defining `effective mass' in crystals using quantum mechanics according to~\cite[x]{27-Wiki}.
To measure the previous relations after the storage vector event occurrence on $1_{x} $ and $1_{y} $, benefiting from relations (4.14a); (4.14b) and (4.18), explicitly,
\vspace{-16pt}
\begin{flushleft}
 $$\therefore \mathop{{\rm \oint }}\nolimits_{A}^{^{} } \mathrm{\mathbf{D}\: d}\mathrm{\mathbf{A}}_{\left(xy,x_{i} y_{i} \right)} =\mathop{{\rm \int }}\nolimits_{}^{^{} } \frac{\Sigma }{\Sigma \to \Sigma ^{2} f\left(t\right)^{-1} } \mathrm{d}\Sigma =\left\{\frac{\Sigma ^{2} \tau _{\Sigma } ^{2} }{1_{x,y} \to 2\left(xy,x_{i} y_{i} \right)} \right\} \:\:\:\:\:\:\:\:\:\:\:\:\:\:\:\:\:\:\:\:\:\:\:\:\:\:\:\:\:(4.10)$$
\end{flushleft}

\small
 $$=\left|\frac{e^{2} t^{2} }{2A^{2} \tau ^{2} } \right|_{xy\left|x_{i} y_{i} \right. }^{x_{i} y_{i} \left|xy\right. }=\left|\frac{-e_{1} t_{1} }{\sqrt{2} xy\tau } \right|,\left|\frac{-e'_{2} t_{2} }{\sqrt{2} x_{2} y_{2} \tau } \right|\:\:\:\:\:\:\:\:\:\:\:\:\:\:\:\:\:\:\:\:\:\:\:\:\:\:\:\:\:\:\:\:\:\:\:\:\:\:\:\:\:\:\:\:\:\:\:\:\:\:\:\:\:\:\:\:\:\:\:\:\:\:\:\:\:\:\:\:\:\:\:\:\:\:\:\:\:\:\:\:\:\:\:\:\:\:\:$$ $$\equiv \frac{e\sqrt{e_{1} e_{2} e'_{2} t_{1} \left|t_{2} t_{1} \right. } }{\sqrt{2} \left(xy\left|x_{2} y_{2} \right. \right)\tau \mathop{{\rm \int }}\nolimits_{P}^{^{} } \mathrm{d}{\rm \jmath}} ,\frac{e\sqrt{e_{1} e_{2} e'_{2} t_{2} \left|t_{1} t_{2} \right. } }{\sqrt{2} \left(x_{2} y_{2} \left|xy\right. \right)\tau \mathop{{\rm \int }}\nolimits_{P}^{^{} } \mathrm{d}{\rm \jmath}} \; \mathrm{in} \; \mathrm{sAm}^{-2} .\:\:\:\:\:\:\:\:\:\:\:\:\:\:\:\:\:\:\:\:\:\:\:\:\:\:\:\:\:\:\:\:\:\:\:\:\:\:\:\:\:\:\:\:\:\:\:\:\:\:\:\:\:\:\:\:\:\:
 \:\:\:\:\:\:\:\:\:\:\:\:\:\:\:\:\:\:\:\:\:\:\:\:\:\:\:\:\:(4.11)$$

\vspace{6pt}
\noindent where discrete valued area of $\mathrm{d}\mathrm{\mathbf{A}}_{\left(xy,x_{i} y_{i} \right)} $ is perpendicular to electric displacement field, \textbf{D} in discrete axes of $e$'s movement between coordinates $xy$ and $x_{i} y_{i} $, or by case in point to the latter, $x_{2} y_{2} $. Discrete representation, $\left|\frac{-e_{1} t_{1} }{xy\tau } \right|,\left|\frac{-e'_{2} t_{2} }{x_{2} y_{2} \tau } \right|$$\cos \left(\frac{\pi }{4} \right)$ or $\sin \left(\frac{\pi }{4} \right)$, measured in sAm$^{-2} $ or, Cm$^{-2} $, states that: Negative stationary time as `stored time' from the `past proper time' was once from either of complex plane integral of $\jmath$, being multiplied into each other from both outcomes via depository vector $\Sigma $.

Let this integral be of the square root resultant for $\tau _{\Sigma } $ via an `OR closure operator', or, $\mathop{\vee }\limits_{i=2}^{n} i$. It is perceived that the remaining negative components in SR relation (4.13), resulting negative planer values by implying limit, $\mathop{\lim }\limits_{P\to -\infty } $, under radical for $P$ of the integral, could be any OR'ed value assigned to $\jmath$ in its summation form, $\sum {\rm \jmath} $, in the negatively progressing definite contour integral result. In other words, ${\rm\jmath }$'s sum by its assigned complex plane function is also independent of a choice of coordinates. This is given after enunciating the following hypothesis relevant to previous relation, complying with future relation (4.18).

\vspace{8pt}
\noindent \textbf{Hypothesis 4.1.} \textit{Two explicit storage outcomes on computable data in a memory cell may exist as charge, $e$, in an n-dimensional space traveling in a bi-reachable path as length contraction/expansion via $\sqrt{xy} $ advanced to the space area and volume occupation. The following proof by deduction, leads to the current hypothesis:} \\

\noindent \textbf{Proof by deduction. }\textit{Two explicit storage outcomes on data for a super-helical memory cell is envisaged to exist as charge, $e$, given in (4.11) ready to be stored integrally, therefore measured in Coulombs per square meter when the depository vector, $\Sigma $, in closure, establishes stored time for either outcome. At the beginning, once the depository vector, $\Sigma $, establishes entangled time t in favour of entangled charge, $e$, equal to total quantum charge $Q$ from (4.18), would therefore be in confined quantum entanglement on area $A$, as, $Q\left|\sqrt{QQ'} \right. \left(2\sqrt{xy} A_{x_{i} y_{i} } \right)^{-1} $. This type of confinement is for outcomes of (4.11) and (4.18), where $\psi \left(\mathbf{r}\right)$, is Schrödinger's wavefunction, and time barrier $t_{1} \left|t_{2} t_{1} \right. $, is the confinement operator on any occurring entanglement. This entangling area of total charge confinement, thus generates a volume of freedom of area of choice $A_{x_{i} y_{i} } $ for charge, $e$, via $\sqrt{Q\sqrt{QQ'} } \left(2\sqrt{xy} {\rm {\mathcal V}}\right)^{-1} $, measured in Coulombs per cube meter multiplied by chosen meter. The $\sqrt{xy} $ dimension is calculable by Lorentz transformations concerning length contraction, hereon the context of charge, $e$'s path bi-reachability as length contraction/expansion.}\textbf{\textit{ }}
\vspace{8pt}
\textbf{}

Imperatively, the conduction of `image logic' might to some people stand as a chronicle point for merely photosensitive materials and devices such as charged coupled devices (CCDs) illustrated in Fig.$\,$1.2, \S\ref{section1}. Nevertheless, `image logic' is a factual grade of computational processes, where one or more than one logic and image state, are combined into one data collection in release (Fig.$\,$1.1, \S\ref{section1}).

In reason, the image logic's output is of permitting a set of phenomena as bodies in motion and their entanglement on loop factors of time themselves in the context of simultaneity. That is, Einstein's synchronization procedure by Einstein in~\cite{13-Einstein}, using Lorentz transformations on observable data path in expansion/contraction (proper length $x'$, when simultaneous paths for the body stands applicable with rise and fall of $y'$). The latter transformation is paradoxical to just, `length contraction' recalled by Nave~\cite{25-Nave}, and time, now into a proper-stored time $\tau _{\Sigma } $, is proportional to proper time in Minkowski's metric for relations (4.10) and (4.11). This proportionality, is due to temporal coordinate time $t$ and $x,y,z$ as orthogonal spatial coordinates, where axis $z$ is too temporal; $t_{1} $ and $t_{2} $ however, are in conjunction with superposition of charge $e$, or to be more specific,
\vspace{-20pt}
\begin{flushleft}
\[\tau _{\Sigma } \propto \tau _{0} \left|{\rm \; if\; \; }x'=x=y{\rm \; ,}\right. \Delta x'=\frac{\Delta x}{\sqrt{1-v_{e,x}^{2} c^{-2} } } \Rightarrow \mathop{\lim }\limits_{y'\to x} \left(y'\ne y\right)\leftrightarrow \mathop{\lim }\limits_{y'\to y} \left(y'=y\right) \]
\end{flushleft}
\vspace{-20pt}
\begin{flushleft}
$\because {\rm \;}z=r\left|f\left(r\right)\mapsto t\right. , $
\end{flushleft}

\noindent where $r$, is the radial distance whose values are mapped to time coordinate $t$ during angular change of $r$ in (4.17) between trap sites with unity to Figs.$\,$4.1 \& 4.2. Hence, $\Delta \tau $ equals to
\vspace{-3pt}
$$\sqrt{{\rm \; }\Delta t^{2} -\left(\frac{v_{e,x}^{} \Delta t}{c} \right)^{2} -\left(\frac{v_{e,y}^{} \Delta t}{c} \right)^{2} -\left\{\begin{array}{l}\left(\frac{v_{e,z}^{} \Delta t}{c} \right)^{2} \stackrel{\mathrm{iff}}{\longrightarrow}\left(\frac{\Delta t,\Delta t\left(x\right)_{xy} }{ct_{x} ,t_{y} } \right)^{2} + \\
\ \ \ \ \ \ \ \ \ \ \ \ \ \ \ \ \ \ \ \left(  \frac{\Delta ty_{xy} \vee y'_{xy} }{ct} \right)^{2} \end{array} \right\}}\: , \ \ (4.12\mathrm{a}) $$

\noindent where $\Delta t\left(x\right)_{xy} $ and $\Delta ty_{xy} $ or to the latter's complement, $\Delta ty'_{xy} $, come from the two engaged fields, one time varying \textbf{B}-field multiplied by its external field, and the other from CNT component across N and S poles of the staircase construction.

\small In continue, since if charge $e$, tends to travel both $x$ and $y$ directions simultaneously (here, being in two places at the same time), and thus sourced in favour of $z$ (here, is due to the travel from nano-thin layers of quantum wells or from lower electrode layers in an induced state to memory cell), charge $e$, should therefore possess a speed twice as much greater than its original speed relative to its speed performed in the memory cell's trap site. \\

\noindent Consequently, after simplifying, we should then elicit,
\vspace{-14pt}
\begin{flushleft}
$$\Delta \tau =\sqrt{{\rm \; }\Delta t^{2} -\left(\frac{v_{e,x} \Delta t}{c} \right)^{2} -\left(\frac{v_{e,y} \Delta t}{c} \right)^{2} -\left\{\left(\frac{v_{e,z} \Delta t}{c} \right)^{2} \stackrel{\mathrm{iff}}{\longrightarrow}\left(\frac{\Delta tx}{ct_{x} } \right)^{2} +\left(\frac{\Delta tx}{ct_{x} } \right)^{2} \right\}} {\rm \; ,}$$\end{flushleft} and when $\left\{x'=x=y\right\}\mapsto f\left(v_{e} \right)\Delta t$, thence,
\vspace{-12pt}
\begin{flushleft}
$$\therefore \sqrt{\Delta t^{2} -\left(\frac{v_{e,x} \Delta t}{c} \right)^{2} -\left(\frac{v_{e,y} \Delta t}{c} \right)^{2} -\left\{2\left(\frac{\Delta tx}{ct_{x} } \right)^{2} \right\}}\; \; \; \; \; \; \; \; \; \; \; \; \; \; \; \; \; \; \; \; \; \; \; \; \; \; \; \; \; \; \; \; \; \; \; \;$$\end{flushleft}
\vspace{-16pt}
\begin{flushleft}
$$=\sqrt{{\rm \; }\Delta t^{2} -\left(\frac{v_{e,x} \Delta t}{c} \right)^{2} -\left(\frac{v_{e,x} \Delta t}{c} \right)^{2} -\left\{2\left(\frac{\Delta tx}{ct_{x} } \right)^{2} \right\}}\; , \; \; \; \; \; \; \; \; \; \; \; \; \; \; \; \; \; \; \; \; \; \; \; \; \; \; \; \; \; \; \; \; \; \; \;$$\end{flushleft}
\begin{flushleft}
\vspace{-12pt}
$$\therefore\Delta \tau =\Delta t\sqrt{{\rm \; 1}-2\left(\frac{v_{e,x} }{c} \right)^{2} -2\left(\frac{v_{e,x} \left|v_{e,xy} \ge c\right. }{c} \right)^{2} } \le \Delta t\sqrt{{\rm \; 1}-\frac{4v_{e,x}^{2} }{c^{2} } } \:\:\:\:\:\:\:\:\:\:\:\:\:\:\:\:\:\:\:\:\:\:\:\:\:\:\:\:\:\:\:\:(4.12{\rm b})$$ \end{flushleft}
\noindent
\vspace{-2pt}
$$\mathrm{where},\; \tau _{\Sigma } \propto \tau _{0} \left|{\rm \; }\Delta \tau =\Delta t\sqrt{{\rm \; 1}-4\left(\frac{v_{e,x} }{c} \right)^{2} -\left(\frac{v_{e,xy} \ge c}{c} \right)^{2} } \right. ,x'=x=y \; \: \mathrm{and \; thus \; the}\; \; \; \; \; \; \; \; \; \; \; \; \; \; \; \; $$ \noindent   {following as},
\vspace{-16pt}
\begin{flushleft}
$$\tau _{\Sigma } \propto \Delta \tau \left|{\rm \; }\Delta \tau =\Delta t\sqrt{{\rm \; }-4\left(\frac{v_{e,x}^{} }{c} \right)^{2} -\left(\mathop{\vee }\limits_{i=2}^{n} i\right)^{2} } \right. =\Delta t\sqrt{{\rm \; }-\frac{4v_{e,x}^{2} }{c^{2} } -4,...,n^{2} _{xy} } =\; \; \; \; \; \; \; \; \; \; \; \; \; \; \; \; $$ \end{flushleft}\vspace{-14pt}
\begin{flushleft}$$ 2\Delta t\sqrt{{\rm \; }-9,...,n^{2} _{xy} -\frac{v_{e,x}^{2} }{c^{2} } } .\:\:\:\:\:\: \:\:\:\:\:\:\:\:\:\:\:\:\:\:\:\:\:\:\:\:\:\:\:\:\:\:\:\:\:\:\:\:\:\:\:\:\:\:\:\:\:\:\:\:\:\:\:\:\:\:\:\:\:\:\:\:\:\:\:\:\:\:\:\:
\:\:\:\:\:\:\:\:\:\:\:\:\:\:\:\:\:\:\:\:\:\:\:\:\:\:\:\:\:\:\:\:\:\:\:\:\:\:\:\:\:\:\:\:\:\:\:\:\:\:\:\:\:\:\:\:\:\:\:\:\:\:\:\:
(4.12{\rm c})$$
\end{flushleft}

\noindent For relative expanded geometric conditions where far memory cell's trap sites are of `tuples or attribute square unit base by location', once $x'=x=y$ is established, for \(\tau _{\Sigma }\) we then acquire,

\bigskip
\vspace{5pt}
\begin{flushleft}
\includegraphics[width=16.8mm, viewport= 0 0 10 25]{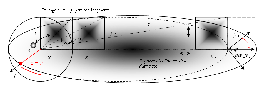}\\
\end{flushleft}
\vspace{-5pt}

\noindent{\footnotesize \textbf{Figure 4.2.} Charge storage film's equipartiton fragments or `trap sites' of the memory cell enables the charge from the highly conductive localized CNT, to be stored. The induced charges in entanglement are also being stored once the quantized energy of the thin ONO film is associated with a nano-scale CNT channel. It is determined that this thin film has a quasi-quantized energy state due to the stored charges having a quantized voltage increment value as claimed by Choi \textit{et al.} in~\cite{06-Choi}. This illustration also explains the paradoxical deportment of the storable data on an expandable data path meeting conditions of simultaneity on \textbf{$x_{i} ,y_{i} $} of \textbf{$x',y'$}, solely when deductions of (4.12) are dealt with.
\vspace{-12pt}
\small
\begin{flushleft}
$$\because \Delta \tau =\frac{2\Delta t}{c} \sqrt{-v_{e,x}^{2} -9,...,n^{2} c^{2} _{xy} } \le \left\{\left(t_{1} \left|t_{2} \right. \right)\cdot 2\Delta t\mathop{{\rm \int }}\nolimits_{P}^{^{} } \mathrm{d}{\rm \jmath}=\frac{\Delta \tau }{2\mathop{{\rm \int }}\nolimits_{P}^{^{} } \mathrm{d}{\rm \jmath}} \cdot \left(t_{2} \left|t_{1} \right. \right)\right\}\lower3pt\hbox{\rlap{$\scriptscriptstyle\sim$}}< (t_{{\rm E} }^{2} $$\end{flushleft}
\begin{flushright}$=\left(\frac{t_{1} +t_{2} }{\sqrt{2} {\rm }} \right)^{2}),$\end{flushright}\vspace{-12pt}
$$\therefore \tau _{\Sigma }^{2} =\Delta t^{2} \mathop{\lim }\limits_{P\to -\infty } f\left(P\right)\stackrel{\sum {\rm } }{\longrightarrow}\mathop{{\rm \int }}\nolimits_{P}^{^{} } \mathrm{d}{\rm \jmath}^{-1} , \; \mathrm{thus},\:\:\:\:\:\:\:\:\:\:\:\:\:\:\:\:\:\:\:\:\:\:\:\:\:\:\:\:\:\:\:\:\:\:\:\:\:\:\:\:\:\:\:\:\:\:\:\:\:
\:\:\:\:\:\:\:\:\:\:\:\:\:\:\:\:\:\:\:\:\:\:\:\:\:\:\:\:\:\:\:\:\:\:\:\:\:\:\:\:\:\:\:\:\:\:\:(4.13)$$

$$\therefore \tau _{\Sigma }^{2} =\frac{\Delta t^{2} }{2\Delta \tau \left(r\sum {\rm \jmath} \right)} =\frac{\left(t_{1} \left|t_{2} \right. \right)\dot{t}_{0} }{2\left(r\sum {\rm \jmath} \right)} =\frac{1\left(t_{1} \left|t_{2} \right. \right),1\left(t_{2} \left|t_{1} \right. \right)}{\mathrm{d}r\mapsto t} ,\:\:\:\:\:\:\:\:\:\:\:\:\:\:\:\:\:\:\:\:\:\:\:\:\:\:\:\:\:\:\:\:\:\:\:\:\:\:\:\:\:\:\:\:\:\:\:\:\:\:\:\:\:\:\:\:\:\:\:
\:\:\:\:\:(4.14{\rm a})$$
$$\therefore \tau _{\Sigma } =\frac{-\sqrt{t_{1} \left|t_{2} t_{1} \right. } }{\sqrt{2} \tau } ,\frac{-\sqrt{t_{2} \left|t_{1} t_{2} \right. } }{\sqrt{2} \tau } =1\left(t_{1} \left|t_{2} \right. \right),1\left(t_{2} \left|t_{1} \right. \right)=\left\{\begin{array}{l} {{\rm \; }\left|\dot{t}_{0} \right|\cos \left({\tfrac{\pi }{4}} \right){\rm \; }} \\ {{\rm \; }\left|\dot{t}_{0} \right|\sin \left({\tfrac{\pi }{4}} \right)} \end{array}\right\}\approx \; \; \; \; \; \; \; \; \; \; \; \; \; \; $$ $0.7071\left|\dot{t}_{0} \right|\equiv \tau _{0} ,\:\:\:\:\:\:\:\:\:\:\:\:\:\:\:\:\:\:\:\:\:\:\:\:\:\:\:\:\:\:\:\:\:\:\:\:\:\:\:\:\:\:\:\:\:\:\:\:\:\:\:\:\:\:\:\:\:\:\:
\:\:\:\:\:\:\:\:\:\:\:\:\:\:\:\:\:\:\:\:\:\:\:\:\:\:\:\:\:\:\:\:\:\:\:\:\:\:\:\:\:\:\:\:\:\:\:\:\:\:\:\:\:\:\:\:\:\:\:\:\:
\:\:\:\:\: \;(4.14{\rm b})$
\vspace{8pt}\\
\noindent or, equal to symmetric time in rest in the context of time depository in simultaneity. By definition,
\vspace{-1pt}
\[\therefore \tau _{\Sigma } =\frac{{\rm stationary\; time\; factor\; of\; past\; proper\; time\; data}}{{\rm current\; proper\; time\; data\times radial\; time\; distance\; of\; complex\; plane\; integral\; }} \equiv \tau _{0}  ,\; \; \; \; \; \; \; \; \; \; \; \; \;\]
\begin{flushleft}
         $\left. \varphi =\arg \left(z\right){\rm \; },{\rm \; }r=\sqrt{x^{2} +y^{2} } ,{\rm \; }x=r\cos \varphi {\rm \; },{\rm \; }y=r\sin \varphi \right\}$.$\:\:\:\:\:\:\:\:\:\:\:\:\:\:\:\:\:\:\:\:\:\:\:\:\:\:\:\:\:\:\:\:\:\:\:\:\:\:\:\:\:\:\:\:\:\:\:\: \; \; \; \; \; \; \; \; \; \; \; \; \; \; (4.15)$
\end{flushleft}

\noindent An example of stored radar pulses is by time of events, $t_{{\rm E} } $ in (4.13), originated from Wright~\cite{41-Wright}'s discussions on relativity, now in favour of (4.15). Therefore, associating radial time distance $r$ for a vector valued area of $\mathrm{d}\mathrm{\mathbf{A}}_{xy} $ which is perpendicular to electric displacement field \textbf{D}, in terms of radial area $\mathrm{d}A_{r} $, generates an electromotive force (emf), ${\rm {\mathcal E}}$ for (4.16), in a certain time interval divided by resistance $R$ for the selective trap site area $A$. This selection of area occurs during field's displacement, since a time varying {\textbf B}-field system on charge $e$ for a closed loop capacitance is established, performing $e\stackrel{L_{V} ,{\rm \; }H_{V} }{\longrightarrow}{\left| \Psi  \right\rangle} $, as a free discrete structure of stored charges with respect to storable passing time $\tau $, interdependent with time phase $\phi _{2} $, in regard to $WR.\phi _{1} :RD.\phi _{1} $. This discrete structure urges to recall (1.6), \S\ref{section1}, as an electromagnetic relation defining circuit's energy expectation. The behaviour derived from Maxwell's equations, is expressed as follows

\vspace{-6pt}
$$\therefore \mathop{{\rm \oint }}\nolimits_{C}^{^{} } \mathrm{\mathbf{E}}\mathrm \: {\mathrm{d}}\mathrm{\mathbf{I}}\wedge A_{r} =\frac{-\mathrm{d}\Phi _{m} }{\mathrm{d}t} \stackrel{\nabla \cdot \mathrm{\mathbf{B}}\bigcap A_{r} }{\longrightarrow}\left|\mathrm{\mathbf{D}}\right|=\frac{\mathrm{d}QA_{r} }{\mathrm{d}A_{xy} \mapsto f_{A} } \equiv \frac{{\rm {\mathcal E}}t}{RA}  ,                                            \:\:\:\:\:\:\:\:\:\:\:\:\:\:\:\:\:\:\:\:\:\:\:\:\:\:\:\:\:\:\:\:\:\:\:\:\:\:\:\:\:
\:\:\:\:\:\:\:\:\:\:\:\:\:\:\:\:\:\:\:\:\:\:\:\:\:\:\:\:\:\:\:\:\:\:\:\:\:\:\:\:\:\:\:\:\:\:\:\:\:\:\:\: (4.16)$$
\vspace{-2pt}

\noindent where \textbf{E}, is the electric field, d\textbf{I}, is the infinitesimal element of contour $C$. In the denominator, after implying $\nabla \cdot \mathrm{\mathbf{B}}=A_{r} $, area $A_{xy} $ is treated as a derivative of area function $f_{A} $, denoting the area with radial distance $A_{r} $ promoted to some other 2D-trap site with index $x_{i} y_{i} $, otherwise remaining as $xy$ for electric displacement field \textbf{D}. Therefore, obtaining an asynchronous contour integral whilst in area-volume's geometric state of integration to the point of Ampere's Circuital law of capacitance extracted from Maxwell's 1861 paper~\cite{23-Maxwell}; and~\cite[viii]{27-Wiki}, is thus plausible; or,

\vspace{-6pt}
$$\therefore \mathop{{\rm \int \oint }}\nolimits_{A}^{^{} } \mathrm{\mathbf{D}} \: \mathrm{d}\mathrm{\mathbf{A}}_{xy} \mathrm{d}A_{r} =QA_{xy}^{-1} \: \mathrm{d}A\mapsto A_{r\to x_{i} y_{i} }^{} =QA_{xy}^{-1} A_{r\to x_{i} y_{i} }^{-1} J \: \mathrm{d}r\mathrm{d}\varphi =QA^{-2} r \; \mathrm{d}r\mathrm{d}\varphi {\rm \; },$$      \begin{flushright}(4.17)\end{flushright}
\vspace{-13mm}
\begin{flushleft}
$$\mathrm{where},\; J=\det \frac{\partial \left(x,y\right)}{\partial \left(r,\varphi \right)} =\left[\begin{array}{cc} {{\tfrac{\partial x}{\partial r}} } & {{\tfrac{\partial x}{\partial \varphi }} } \\ {{\tfrac{\partial y}{\partial r}} } & {{\tfrac{\partial y}{\partial \varphi }} } \end{array}\right]=r\cos ^{2} \varphi +r\sin ^{2} \varphi =r, \; \mathrm{explicitly}, \; \;  \; \;  \; \;  \; \;  \; \;  \; \;  \; \; $$
\vspace{-1mm}
$$\therefore \frac{\left|\mathrm{\mathbf{D}}\right|}{1\leftrightarrow \left|\mathrm{\mathbf{A}}_{yx} \right|} =\left|\frac{QA}{2A^{2} } \right|\; \;  \; \;  \; \;  \; \;  \; \;  \; \;  \; \; \; \;  \; \;  \; \;  \; \;  \; \;  \; \;  \; \; \; \;  \; \;  \; \;  \; \;  \; \;  \; \;  \; \; \; \;  \; \;  \; \;  \; \;  \; \;  \; \;  \; \; \; \;  \; \;  \; \;  \; \;  \; \;  \; \;  \; \; \; \;  \; \;  \; \;  \; \;  \; \;  \; \;  \; \; \; \;  \; \;  \; \;  \; \;  \; \;  \; \;  \; \; \; \;  \; \;  \; \;  \; \;  \; \;  \; \;  \; \; $$
\vspace{-1mm}
$$\leftrightarrow \mathop{\mathop{{\rm \int }}\limits_{{\rm {\mathcal V}}} }\nolimits_{}^{^{} } \rho _{e} \left(\mathrm{\mathbf{r}}\right)\: \mathrm{d}{\rm {\mathcal V}}=\frac{Q_{free\to bailed\leftarrow free} }{2A_{freedom} } \equiv e\mathop{{\rm \int }}\nolimits_{}^{^{} } \left|\psi \left(\mathrm{\mathbf{r}}\right)\right|^{2} \: \mathrm{d}\mathrm{\mathbf{r}}=\frac{Q\left|\sqrt{QQ'} \right. }{2\sqrt{xy} A_{x_{i} y_{i} } } {\rm =\; }\frac{Q}{2{\rm {\mathcal V}}}, \:\:\:\:\:\:\:\:\:\:\:\:\:\:\:\:\:\:\:\:(4.18)$$
\end{flushleft}

\noindent where this therefore implies to $Q\left(t_{2} \right)\mapsto Q\left(t_{2} \left|t_{1} \right. \right)\prec WR.\phi _{1} :RD.\phi _{1} \succ $ brought by subsequent relations, (4.19) and (4.20), and the $\therefore \frac{\left|\mathrm{\mathbf{D}}\right|}{1\leftrightarrow \left|\mathrm{\mathbf{A}}_{yx} \right|} $'s righthand result in (4.18). The latter deduction in (4.18), is the charge density on the quantum capacitor's chosen plate via radial distance $r$ between $xy$ and  $x_{i} y_{i} $, removing the problem of blocking flows of electric current due to the intrinsic nature of capacitors on charge's trap sites. Quantum charge density $\rho _{e} \left(\mathrm{\mathbf{r}}\right)$, in the same relation, represents continuous charge density of total quantum charge, $Q=e\mathop{{\rm \smallint }}\nolimits_{}^{^{} } \left|\psi \left(\mathrm{\mathbf{r}}\right)\right|^{2} \: \mathrm{d}r$, over a volume ${\rm {\mathcal V}}$, which could be either spherical or elliptical of the region (observe, Fig.$\,$4.2). This regional volume conserving $Q$ transformations, displays a readable displacing charge, $Q_{free\to new{\rm \; }bailed{\rm \; }on{\rm \; }display} $, converted into quantum charge density via wavefunction $\psi \left(\mathrm{\mathbf{r}}\right)$.

The momentum of the storable charge $e$ as $mv_{e} $, also inflicts upon radial time change of (4.15), as De Broglie's wavelength representation of relativistic momentum, $p=\gamma m_{0} v_{e} $ by Nave~\cite{25-Nave}, convertible to a photon momentum, $h\lambda ^{-1} $, as too, storable from light emitting conditions (e.g., ELEDs). The light emitting components, if used, are for representing photosensitive image-logic states, where mass in this case is stretched out with relevance to (1.4), \S\ref{section1}, between all traps on charge $e$, as new $mv_{e} v_{e'} $ $\in \prod E_{e\stackrel{L_{V} ,{\rm \; }H_{V} }{\longrightarrow}{\left| \Psi  \right\rangle} }  $, until the moment of charge deposit.

With relevance to the previous paragraph, the graph of the evolution of $e$'s mass is shown in Fig.$\,$4.1 with the involved Gaussian curvature-states signifying the finite world of charge $e$, in the course of radial time. Tab.$\,$4.1, also represents the evolution states of $e$'s mass in form of formulae. Generally, radial distance $r$, as radius of time $t$, is from the pole in the polar coordinate system converted from the Cartesian form on angle $\varphi $, defined on the complex plane. It is thus noted that the extraction and travel of charge $e$, in terms of  $e^{2} e_{1} e_{2} $, is with restriction to time via \textbf{B}-field lines (here as the carrier's worldline) of CCNT to the electric lines of transference, between $xy$ and $x_{i} y_{i} $ via $\mathrm{d}A_{r} $ across the fields' flux volume.

In conjunction with (4.17), symbol $\nabla $, represents the del operator from the divergence theorem, where the $\nabla \cdot \mathrm{\mathbf{B}}=0$ representing no charge in the field's region, assumably tracks down radial distance wherever it ends up for charge $e$. This tracking of distance is defined as either condition of $\Sigma $ in (4.9) between $xy$ and $x_{i} y_{i} $ traps representing $\Sigma ^{2} $ from (4.8), implied via compositional intersection $\nabla \cdot \mathrm{\mathbf{B}}\bigcap $$A_{r} $ from (4.17). Also, destination area $A$ in (4.16) and (4.17), is given as an area-element in the Cartesian coordinates, defined as an infinitesimal area relating to quantum dot before Cartesian conversion. Symbol $J$ of (4.17), represents the Jacobean determinant of the coordinate conversion formula by (4.19) and (4.20). Charge $e$, is permitted to choose by freedom of choice via radial area $A_{r} $, where composition $A_{xy} A_{r\to x_{i} y_{i} } $ is stipulated according to the mapping, $\mathrm{d}A\mapsto A_{r\to x_{i} y_{i} }^{} $, of (4.17). This elementary area eventually expands into the memory cell's storage limit with reason to relations (4.17) and (4.16), ensuing (4.18) accordingly.

This is why semantically, choosing  $Q_{bailed} $ as a bailed quantity (provisionally $Q_{enclosed} $ in Gauss's law context) of electric charge $e$, becomes relevant to the data transmission operation. The total charge ($Q$) described as a `bailed quantity', is a quantity of electric charge; a free charge provisionally remains free until the moment of trap. Let the opposing form be $Q_{bailed} ^{{'} } $, the chosen $Q_{bailed} $ would then be expressed into a principle volume of charge conservation in terms of

\vspace{1pt}
\begin{flushleft}
$Q\left(t_{2} \right)=\left(t_{1} \right)+Q_{in} -Q_{out} ,{\rm \; }\mathrm{if}{\rm \; }Q\left(t_{2} \right)\mapsto \left(t_{2} \left|t_{1} \right. \right){\rm \; thus,\; }\therefore Q\left(t_{2} \left|t_{1} \right. \right)=Q\left(t_{1} \left|t_{2} \right. \right)+Q_{in} -Q_{out} $,$\:\:\:\:\:\:\:\:\:\:\:\:\:\:\:\:\:\:\:\:\:\:\:\:\:\:\:\:\:\:\:\:\:\:\:\:\:\:\:\:\:\:\:\:\:\:\:\:\:\:\:\:\:\:\:\:
\:\:\:\:\:\:\:\:\:\:\:\:\:\:\:\:\:\:\:\:\:\:\:\:\:\:\:\:\:\:\:\:\:\:\:\:\:\:\:\:\:\:\:\:\:\:\:\:\:\:\:\:\:\:\:\:\:\:\:(4.19)$

$t_{1} \underline{\succ }WR.\phi _{1} \stackrel{}{\longrightarrow}Q_{bailed} :RD.\phi _{1} \stackrel{}{\longrightarrow}Q_{bailed} ^{{'} } \underline{\prec }t_{2} $ , $\therefore t_{2} \underline{\succ }RD.\phi _{1} \stackrel{}{\longrightarrow}Q_{bailed} ^{{'} } \equiv RD.\phi _{1} \stackrel{}{\longrightarrow}Q_{free\leftarrow bailed\to free} \equiv RD.\phi _{1} \stackrel{}{\longrightarrow}Q_{free\to new{\rm \; }bailed{\rm \; }on{\rm \; }display} $.$\:\:\:\:\:\:\:\:\:\:\:\:\:\:\:\:\:\:\:\:\:\:\:\:\:\:\:\:\:\:\:\:\:\:\:\:\:\:\:\:\:\:\:\:\:\:\:\:\:\:\:\:\:\:\:\:\:
\:\:\:\:\:\:\:\:\:\:\:\:\:\:\:\:\:\:\:\:\:\:\:\:\:\:\:\:\:\:\:\:\:\:\:\:\:\:\:\:\:\:\:\:\:\:\:(4.20)$
\end{flushleft}

\noindent Discrete mathematical symbols, `precedes from', $\prec $, and `succeeds from', $\succ $, as time predecessors of write and read time phase $\phi _{1} $ on charge density $Q\left(t_{2} \right)$ and $Q\left(t_{2} \left|t_{1} \right. \right)$, are derived from (4.19) implied to $Q_{bailed} $ and $Q_{bailed} ^{{'} } $ respectively, which is well-defined in data read/write operations. The first part of (4.19), denotes the principle of charge conservation~\cite[v]{27-Wiki}; the second part however, as mapped, denotes the symmetric problem of a quantity of the elementary charge $e$, flowing into as $Q_{in} $ and flowing out of the volume as $Q_{out} $ between entangled time $\left(t_{1} \left|t_{2} \right. \right)$ and $\left(t_{2} \left|t_{1} \right. \right)$, simultaneously. The $\left(t_{1} \left|t_{2} \right. \right)$ component in (4.19) and other conceiving relations from the past are `highly symmetric', where the symmetry itself appreciates Noether's Theorem explained by Baez in~\cite{01-Baez}; and~\cite[vi]{27-Wiki}. Reasoning that in this case for the time symmetry, the time variable is said to be at any locus to its increment/decrement position, thus entangled between at least 5 dimensions over a one-dimensional manifold (time) out of an N-dimensional universe. Hence $\left(t_{1} \left|t_{2} \right. \right)$ or its symmetry $\left(t_{2} \left|t_{1} \right. \right)$, `commutatively' in logic, represents stationary time factor $\dot{t}_{0} $-dependant whilst being of proper time $\tau $-dependant, as the stored past or future proper time $\tau _{0} $. Therefore, this duality of time in behaviour, advocates the reality of time factor $\dot{t}$, paradoxically in form of $\dot{t}^{2} \ne \dot{t}^{2} $ from past relations (4.5) and (4.14b) pertinent to statements on the subject of Russell's Paradox (e.g.,~\cite{31-Quine};~\cite{34-Russell}). This `time at rest symmetry' is in the dilated magnetic moment $\mu $ of charge $e$, read and write time phase $\phi _{1} $, via evaluating $\phi _{2} $'s feedback for the stored data-bit process (for more details, refer to \S\ref{section5}). Bear in mind, this $\left(t_{1} \left|t_{2} \right. \right)$ and its symmetry $\left(t_{2} \left|t_{1} \right. \right)$ concept, is of

\vspace{6pt}
\noindent \textbf{Theorem 4.1. }\textit{Let time barrier or restriction $\left(t_{1} \left|t_{2} \right. \right)$, and its symmetry stand never an anomaly, since this method of representing time, is itself merely a `presentation of time continuity equation' as a whole, which is constantly interconnected or intact to its past, current and future dilation. In satisfaction of this theorem, the following conditions would rely as equivalent:} \\ \vspace{-12pt}

\begin{enumerate}
\item [{(1)}]\textit{The interconnectivity of time dilation, is independent to event occurrences whilst being dependent symmetrically to its proper (as proper time $\tau $) and its rest, simultaneously.}

\item [{(2)}]\textit{The `rest time' given in condition (1) of this theorem, could be of $t_{1} $ or $t_{1} t_{2} $, as connecting past-to-future and the latter definition's symmetry $t_{2} t_{1} $, connecting future-to-past, correspondingly.}  \textbf{}
\vspace{6pt}
\end{enumerate}

\noindent Relations (4.12) to (4.20), do benefit from the basics of Einstein's twin paradox in SR. Relevantly, say for instance, we imagine charge $e$ as `stationary twin' relative to its `travelling twin' $e_{2} e'_{2} $ in Minkowski's diagram~\cite{33-Rindler}. Now for the $e$'s travelling twin, envisage it to be confined into `plane of simultaneity' via one space-time coordinate representing the occurred superposition, converting superposition data-length $x_{\Delta }^{2} $, into timelike scenario from the stored data's past events. The past events indicate a spacelike scenario for charge $e$ due to `minimum velocity favouring a dimensional-course of entanglement' in quantum scales of quantity, or,
\vspace{-1pt}
\begin{flushleft}
$\forall x_{\Delta }^{2} \in \sum ^{2} \left|x_{\Delta } \right. \equiv$ $$ \frac{f\left(x_{\Delta 0} \right)}{\mathop{\lim }\limits_{\tau _{\Sigma } \to \infty } v{\rm \; }\mathbf{o}\left|\tau _{\Sigma } \right|} =\frac{v_{e}^{} t_{\left(x_{i} y_{i} ,xy\right)\leftrightarrow z} }{t_{z\leftrightarrow \left(xy,x_{i} y_{i} \right)} \sqrt{{\rm \; }\frac{-4v_{e}^{2} -c_{z\leftrightarrow \left(xy,x_{i} y_{i} \right)}^{2} }{v_{e}^{2} \left|c^{2} \right. } } } =\frac{v_{e}^{2} t_{1} t_{2} }{2\tau \left|v_{e,z\leftrightarrow \left(xy,x_{i} y_{i} \right)} \right|\lower3pt\hbox{\rlap{$\scriptscriptstyle\sim$}}>c}$$ $\; \; \; \; \; \; \; \; \; =x_{\Delta } \mathbf{o}\sum ^{2} ,$ where,\vspace{-6pt} $$\:{\rm \; }c_{z\leftrightarrow \left(xy,x_{i} y_{i} \right)}^{} \buildrel\wedge\over= v_{e,z\leftrightarrow \left(xy,x_{i} y_{i} \right)}^{} \buildrel\wedge\over= v_{e}^{} {\rm \; },{\rm \; }\lambda =\frac{x_{\Delta 0} x_{\Delta } }{x_{\left(xy,x_{i} y_{i} \right)} } =n\cdot p=\frac{p^{2} }{2\Delta p} ,{\rm \; }n\in {\mathbb R}^{3\mapsto 2n} \; \; \; \; \;  (4.21)$$\end{flushleft}

\noindent satisfying conditions of $v>c$ from foundations of SR, expectably (e.g.,~\cite{02-Bassett}). These conditions occur solely, once the velocity factor is taken out of radical expression in the denominator shown in (4.21).

This relation appreciates Einstein's modified first law i.e. `no matter can travel faster than the speed of light' wherein this case, charge $e$ is not travelling faster than $c$, in fact, the entangled coordinate systems' implication via axis ratio $z\leftrightarrow \left(xy,x_{i} y_{i} \right):\left(x_{i} y_{i} ,xy\right)\leftrightarrow z$, folds time itself giving out a spacelike problem indeed! Compare this relation with image logic on Moiré pattern as an interferometric approach~\cite{27-Wiki} which describes the limits of pale and dark zones of pattern for the superimposed lines.

Furthermore, imagine two misaligned patterns geometrically possess a $x_{\Delta 0}^{} \times x_{\Delta }^{} $ non-relativistic-relativistic distance or $x_{\Delta }^{2} $, from the point of hot electron emission per complete pattern distance $x_{\left(xy,x_{i} y_{i} \right)} $, where electrons are bombarded on screen upon either pattern $p$ of $\Delta p$, in coordinates time $t_{\left(xy,x_{i} y_{i} \right)\leftrightarrow z} $. It is thus conceivable that travelling back and forth to either generated carriers' path (the modulus usage is for this reason only), forming aligned or even the same pattern's initial result, should thereby obtain wavelength $\lambda $-result in (4.21). The  resultant of this pattern could be mapped into the characteristics of the ONO film representing quantized energy states of the storage system and employed to develop optoelectronic property.

Notation ${\mathbb R}^{3\mapsto 2n} $ in (4.21), represents the subspace notion of Darboux's Theorem commencing with a Euclidian space geometry to a range of `2\textit{n}-dimensional vector space configuration' upon the involved carriers' parametric properties such as, pattern recognition, travelled distance(s) and their entangled loci as well.

Accordingly, the symbolic choice of `$\buildrel\wedge\over= $' by definition `stands for\dots ', is by Dickinson \& Goodwin of~\cite{12-Dickinson} on issues of schema calculus in discrete mathematics. Hence, for one the discrete property of (4.21) conceives and therefore propagates logic upon the outputs no matter the constraints of this expectation. The discrete mathematical concept on these logic outputs, shall institute the condition of true realistic outcome frame representing the only plausible resultant as a `cavity system' and some `display system' conducting all the factual states of reality between `Boolean logic and image-logic states'. Ergo,

\vspace{6pt}
\noindent\textbf{Hypothesis 4.2.} \textit{`Image logic' is in manifestation as an optical course of absorption-spontaneous emission versus stimulated emission cycle manifestation, once the remaining classical binary and quantum factors of memory cells serve charge $e$, with wavelength $\lambda $, proper to its pattern state given in Eq.$\,$ (4.21). The following proof by deduction, leads to the current hypothesis:} \\  \vspace{-6pt}
\\
\noindent \textbf{Proof by deduction.} \textit{This pattern state of Eq.$\,$(4.21) emerges an image of the charge's confinement, the uncertainty at the point of superposition and thus, a depository vector point of closure confining now past, current and future of that same charge in $xy$ cellular dimensions, or its neighbouring $x_{i} y_{i} $ cellular dimensions. Selective mass ratio of charge $e$, in $\mathbf{B}$-field for its periodic energy pattern via Tab.$\,$4.1, is measured with the Zeeman effect. This effect is analogous to the photon type of $e$'s excited atomic state such as an optical course of absorption-spontaneous emission versus stimulated emission cycle.     }\textbf{\textit{}}
\vspace{2pt}

\section{The chip's compatibility and memory cell's time and data representation}
\markboth{}{The chip's compatibility}
\label{section5}
\vspace{6pt}
Based upon Fig.$\,$2.1\emph{d}, \S\ref{section2}, design and logic, by changing the sheet resistance, $R_{s} $ value, for source and drain electrodes in either pole of the memory system, the p.d. circuitry distribution thereby generates a compatible technology to other FET transistors (or, FET family). Most appropriately, DCFL logic would be compatible in board's layout, subsequent to the chip's fabrication process in some simulating `circuit maker' environment. Potency of data input against data output in this circuit requires registers incorporated into the circuit's logical representation of its computational data, hence to evaluate time phase $\phi _{1} $ and $\phi _{2} $ values for \textit{WR}.$\phi _{1} $ and \textit{RD}.$\phi _{1} $. Subsequently, it would be possible to verify the evaluation process on refreshing time cycles via  $\phi _{2} $'s feedback for the stored data-bit with respect to the stored qubits from the \emph{waterfall effect} components i.e. CCNTs' memory cells, in practice. The cyclic problem in storing data could be complied within form of, bit ratio $\Re _{bit} $, representing data of any type per time function $f\left(t\right)$, as being refreshed periodically for a time and space-varying data (compare this with `a semi-static RAM cell design and logic' from Pucknell \& Eshraghian of~\cite{30-Pucknell}). In this case, time function $f\left(t\right)$, represents time $t$-of-$\phi _{2} $ or, in abstract algebraic terms of mapping and mappability $t{\rm \; }\mathbf{o}\: \phi _{2} $, devoted to ratio bit $\Re _{bit} $, and bit frequency $\nu _{bit} $, where the latter is measured in Hertz accordingly. In other words,
\vspace{-15pt}
\begin{flushleft}
\[\nu _{bit} \equiv \frac{\Re _{bit} }{t} =\frac{in_{bit} }{in_{qbit} \times f\left(t\right)} =\frac{\forall A\in \left\{0,1\right\}}{\forall B\in \left\{\left(\partial \cdot B+\bar{\partial }\cdot B\right),\Delta B\left(\tilde{\partial }\cdot B\right)\right\}\times t{\rm \; }\mathbf{o}\:\phi _{2} } \:\:\:\:\:\:\:\:\:\:\:\:\:\:\:\:\:\:\:\:\:\:\:\:\:\:\:\:\:\:\:\:\:\:\:\:\:\:\:\:\:
\:\:\:\:\:\:\:\:\:\:\:\:\:\:\:\:\:\:\:\:\:\:\:\:\:\:\:\:\:\:\:\:\:\:\:\:\:\:\:\:\:\:\:\:\:\:\:\:\:\:\:\:\:\:\:\:\:\:\:\:\:\:
\:\:\:\:\:\:\:\:\:\:\:\:\:\:\:\:\:\:\:\:\:\]
\vspace{-6pt}
       $$=\frac{\left\{0,1\right\}}{\left\{\left({\rm 0\; 1}\right),\left({\rm 1\; 0}\right),\left({\rm 1\; 1}\right),\left({\rm 0\; 0}\right)\right\}\times t{\rm \; }} ,\:\:\:\:\:\:\:\:\:\:\:\:\:\:\:\:\:\:\:\:\:\:\:\:\:\:\:\:\:\:\:\:\:\:\:\:\:\:\:\:\:
\:\:\:\:\:\:\:\:\:\:\:\:\:\:\:\:\:\:\:\:\:\:\:\:\:\:\:\:\:\:\:\:\:\:\:\:\:\:\:\:\:\:\:\:\:\:\:\:\:\:\:\:\:\:\:\:\:\:\:\:\:\:
\:\:\:\:\:\:\:\:\:\:\:\:\:\:\:\:\:\:\:\:\:\:\:\:\:\:\:\:\:\:\:\:\:\:\:\:\:\:\:\:\:\:\:\:\:\:\:\:\:\:\:\:\:\:\:\:\:\:\:\:\:\:
\:\:\:\:\:\:\:\:\:\:\:\:\:\:\:\:\:\:\:\:\:(5.1)$$
\end{flushleft}
\vspace{6pt}
\noindent and by methods of simplification,
\vspace{-12pt}
\begin{flushleft}
$$\therefore \nu _{bit} \equiv \frac{A_{in} }{B_{in} t} \left|\left. \mathrm{for} \; A_{in} \left\{\begin{array}{l} {0\buildrel\wedge\over= L_{\pm V} } \\ {1\buildrel\wedge\over= H_{\pm V} } \end{array}\right. \right\}\mathrm{for}{\rm \; }B_{in} :{\left| \Psi  \right\rangle} =\prod \limits _{m=1}^{n}{\left| \Psi  \right\rangle} _{m}  \right. , \; m=1,2,...,n \;, \; \mathrm{thus},$$

$\left({\rm 0\; 1}\right)\equiv \left(\mathop{\lim }\limits_{\forall e{\rm \; }\in 0V\to 0} 0.0...\times e,\mathop{\lim }\limits_{\forall e{\rm \; }\in V_{S} \to \infty } e\right)\in \left(L_{\pm V} {\rm \; }H_{\pm V} \right)\; ,\:\:\:\:\:\:\:\:\:\:\:\:\:\:\:\:\:\:\:\:\:\:\:\:\:\:\:\:\:\:\:\:\:\:\:\:\:\:\:\:\:
\:\:\:\:\:\:\:\:\:\:\:\:\:\:\:\:\:\:\:\:\:\:\:\:\:\:\:\:\:\:\:\:\:\:(5.2)$ \end{flushleft}
\begin{flushleft}
\vspace{-8pt}
$$\therefore \left({\rm 0\; 1}\right)\equiv \left(\sum 0.0...\times e ,\sum e \right)=\left(0V{\rm \; }V_{S} \right):{\left| \psi  \right\rangle} =a{\left| 0 \right\rangle} +b{\left| 1 \right\rangle} , \:\:\:\:\:\:\:\:\:\:\:\:\:\:\:\:\:\:\:\:\:\:\:\:\:\:\:\:\:\:\:\:\:\:\:\:\:\:\:\:\:
\:\:\:\:\:\:\:\:\:\:\:\:\:\:\:\:\:\:\:\:\:\:\:\:\: (5.3)$$

where $a$ and $b$, are complex numbers such that, $1=\sqrt{{\rm \; }\left|a\right|^{2} +\left|b\right|^{2} } .\:\:\:\:\:\:\:\:\:\:\:\:\:\:\:\:\:\:\:\:\:\:\:\:\:\:\:\:\:\:\:\:\:\:\:\:(5.4)$
\end{flushleft}
\vspace{-1pt}

\noindent Wavefunction ${\left| \psi  \right\rangle} $ representing pure qubit states could extend, once the recording by \textit{n-}qubit quantum register, the state of $B_{in} $ registers between the chip's subsystems establishes, which therefore requires 2$^{\textit{n}}$ complex numbers, accordingly (e.g. Ref.~\cite{11-D. P. Di Vincenzo}).

In relation (5.3), $V_{S} $ is the high level logic and $0V$, is the low level logic, whereas both states appreciate differential signaling, here as the ends of the same logic defining one qubit of information in quantum computing for signal $B_{in} $; supposedly, the superposition of all the classically allowed logic states like $A_{in} $ now being for $B_{in} $ in wavefunction ${\left| \psi  \right\rangle} $, to be as follows:

\vspace{6pt}
\noindent --------------- Let index notation $\pm V$, in either (5.2) or (5.3), represent a typical applied signal in favour of high and low logic states. By obvious reasons, $\pm V$ would specify whether the conventional current $I$, for the potential difference occurrence is of either direction to the opposite flow direction of electrons multiplied by resistance $R$, derivable from Ohm's law $I=VR^{-1} $. It is then with resemblance to (5.3), in ratio to ${\left| \psi  \right\rangle} $, for the remaining state outcomes on signal $B_{in} $, states $\left({\rm 1\; 0}\right)$, $\left({\rm 0\; 0}\right)$ and $\left({\rm 1\; 1}\right)$ as $\left(V_{S} {\rm \; }0V\right):{\left| \psi  \right\rangle} $, $\left(0V{\rm \; }0V\right):{\left| \psi  \right\rangle} $ and $\left(V_{S} {\rm \; }V_{S} \right):{\left| \psi  \right\rangle} $ achieved respectively, which are thereby predicatively conventional.  --------------- \vspace{6pt}

\noindent Bear in mind, low level logic does not necessarily mean the p.d. occurrence to be purely 0 volts, and it was the delimitation factor that formulated the present scenario representing the `quantum tunneling problem' in clarity between all charges for input signal $B_{in} $. An example of quantum tunneling, classically-forbidden energy state, is given in relation (1.4), \S\ref{section1}, recalling events of $\prod E_{e\stackrel{L_{V} ,{\rm \; }H_{V} }{\longrightarrow}{\left| \Psi  \right\rangle} }  $ for $\Sigma $ of relation (4.11), \S\ref{section4}, on charge $e$. Charge $e$ in this case, is with restriction to being entangled with $e_{2} $ as $e_{1} e_{2} $, otherwise, entangled with $e'_{2} $'s symmetry as $e_{2} e'_{2} $. Time $t_{1} $, however, is with restriction to $t_{2} t_{1} $ as \dots $t_{1} \left|t_{2} t_{1} \right. $ in ratio to proper time $\tau $, for depository's space two-dimensional restriction $xy$ to $x_{2} y_{2} $ as $\left(xy\left|x_{2} y_{2} \right. \right)\tau $.

This notion of space-time entanglement for charge $e$ in terms of, $\left(xy\left|x_{2} y_{2} \right. \right)\tau $, could also represent the cell's neighbouring storage frame, whereas for the involved  logic signals of (5.1) and (5.2), are expressed in terms of total charge $Q$ from relations (4.18-4.20) of \S\ref{section4}, or,
\vspace{-18pt}
\begin{flushleft}
$$\forall B_{in} ,A_{in} \in H_{\mathbf{S}} \left|{\rm \; }\mathop{\mathop{{\rm \int }}\limits_{{\rm {\mathcal V}}\to } }\nolimits_{H_{\mathbf{S}} }^{^{} } \rho _{e} \left(\mathbf{r}\right)\mathrm{d}{\rm {\mathcal V}}=\frac{Q}{2{\rm {\mathcal V}}} \right., \mathrm{whereby}, \; \mathrm{from }\;(5.3) \; \mathrm{then}, \:\:\:\:\:\:\:\:\:\:\:\:\:\:\:\:\:\:\:\:\:\:\:\:\:\:\:\:\:\:\:\:\:\:\:\:\:\:\:\:\:\:\:(5.5)$$

$\therefore \left(A_{in} ,B_{in} \right)\equiv \left(Q,Q\left|\sqrt{QQ'} \right. \right):{\left| \psi  \right\rangle} $, and,$\:\:\:\:\:\:\:\:\:\:\:\:\:\:\:\:\:\:\:\:\:\:\:\:\:\:\:\:\:\:\:\:\:\:\:\:\:\:\:\:\:\:\:\:\:\:\:\:\:\:\:\:\:\:\:\:\:\:
\:\:\:\:\: \; \; \; \;  \; \; \; \;  \; \; \; \;  \; \; \;  \; \; \;  \; \; \;  \; \; \;  \; \; \; \; \: \: \: (5.6)$
\vspace{6pt}
$$\therefore \prod E_{e\stackrel{L_{V} ,{\rm \; }H_{V} }{\longrightarrow}{\left| \Psi  \right\rangle} }  \stackrel{_{{\left| \Psi  \right\rangle} } }{\longrightarrow}\left\{\forall B_{in} \in H_{\mathbf{S}} \stackrel{\Re _{bit} \left(in_{bit} \right)^{-1} }{\longrightarrow}in_{qbit} \times f\left(t\right)\right\}, \; \mathrm{such} \; \mathrm{that},\:\:\:\:\:\:\:\:\:\:\:(5.7)$$

$\left(\partial \cdot B,\bar{\partial }\cdot B\right)=\left(\nabla B+\Delta B,\bar{\nabla }B+\bar{\Delta }B\right)$, and,  $$ \tilde{\partial }\cdot B=\nabla B+\bar{\nabla }B+\bar{\Delta }B \; \mathrm{in} \; (5.1)\; \mathrm{for}, \mathop{{\rm \int }}\nolimits_{_{\mathbf{S}} }^{^{} } \phi _{2}\: \mathrm{d}\phi _{2}. \:\:\:\:\:\:\:\:\:\:\:\:\:\:\:\:\:\:\:\:\:\:\:\:\:\:\:\:\:\:\:\:\:\:\:\:\:\:\:\:\:
\:\:\:\:\:\:\:\:\:\:\:\:\:\:\:\:\:\:\:\:\:\:\:\:\:\:\:\:\:\:\:\:\:\:\:\:\:\:\:\:\:\:\:\:\:\:\:\:\:\:\:\:
\:\:\:\:\:\:\:\:\:\:\:\:\:\:\:\:\:\:\:\:\:\:\:\:\:\:\:\:\:\:\:\:\:\:\:\:\:\:\:\:\:\:\:\:\:\:\:\:\:\:\:\: \; \; \; \; \; \; (5.8)$$
\end{flushleft}
\vspace{2pt}
Let with recall to (5.1) and (5.2), input $A_{in} $ in (5.5) and (5.6), represent a logic signal of 0 or 1. For relations (5.5) to (5.9), input $B_{in} $, represents a combinatorial quantum signal from EDL events' logic occurrence introduced by Talantsev of~\cite{36-Talantsev}; where symbol $\bar{\partial }$, represents `no change' or `non-event' in quantum signal; $\partial $, represents any change in the same quantum signal, $B$, accordingly (see also~\cite{30-Pucknell}, \S11). It is noteworthy to mention that, event-driven logic (EDL) prior to the direct coupled FET logic (DCFL) inverters' integration for the bus line at time $t$, is contemplated as the most suitable in describing qubit storage behaviour. This form of inverters' integration represent the change of logic from 0 to 1 and vice versa. The integration also includes the no-change conditions on 0 or 1 logic, explaining whatever the logic is in-between high and low logic states as variant logic, suspended in the \emph{waterfall effect} storage system's scenario.

Bear in mind, the product sequence $0.0...\times e$ in (5.2), would denote that no electron or most conditionally, a few electrons satisfy a `low state logic'. On the one hand, these electrons represent the $x$ coordinate of some testing electronics graph for input signal $B_{in} $'s two dimensional coordinate system of $\left(x{\rm \; }y\right)$ on p.d. `high and low state logic' occurrences. On the other hand, other logical conditions would imply to the remaining coordinates $B_{in} $ against the well-known Boolean logic on signal $A_{in} $, as by propositional restrictions established in (5.2).

In continue, assuming for minimum $\nu _{bit} $ in either pole $\alpha $ or $\beta $, as $\min \nu _{bit} $, and for the expected $\nu _{bit} $ in both poles $\alpha $ and $\beta $, as $\nu _{bit} \left|_{\beta }^{\alpha } \right. $, we then obtain respectively,
 \vspace{-12pt}
\begin{flushleft}
\[
\min \nu _{bit}  = \frac{{1{\mathop{\rm bit}\nolimits} }}{{2{\mathop{\rm bit}\nolimits} _{\left| \Psi  \right\rangle }  \times 10^{ - n} {\mathop{\rm s}\nolimits} }} = 0.5 \times 10^n {\mathop{\rm Hz}\nolimits} \:{\rm  , }\:\:\:\:\:\:\:\:\:\:\:\:\:\:\:\:\:\:\:\:\:\:\:\:\:\:\:\:\:\:\:\:\:\:\:\:\:\:\:\:\:
\:\:\:\:\:\:\:\:\:\:\:\:\:\:\:\:\:\:\:\:\:\:\:\:\:\:\:\:\:\:\:\:\:\:\:\:\:\:\:\:\:\:\:\:\:\:\:\:\:\:\:\:\]
\[\:\nu _{bit} \left| {_\beta ^\alpha  } \right. = \sum\limits_{\scriptstyle i = 1 \hfill \atop
  \scriptstyle j = 2 \hfill}^{\scriptstyle m \ge i \hfill \atop
  \scriptstyle l \ge j \hfill} {\frac{{i{\mathop{\rm bit}\nolimits} }}{{2j{\mathop{\rm bit}\nolimits} _{\left| \Psi  \right\rangle }  \times 10^{ - n} {\mathop{\rm s}\nolimits} }}}  \ge 0.5 \times 10^n {\mathop{\rm Hz}\nolimits}
. \:\:\:\:\:\:\:\:\:\:\:\:\:\:\:\:\:\:\:\:\:\:\:\:\:\:\:\:\:\:\:\:\:\:\:\:\:\:\:\:\:
\:\:\:\:\:\:\:\:\:\:\:\:\:\:\:\:\:\:\:\:\:\:\:\:\:\:\:\:\:\:\:\:\:\:\:\:\:\:\:\:\:\:\:\:\:\:\:\:\:\:\:\: (5.9)\]
\end{flushleft}

\noindent Components $2\mathrm{bit}_{{\left| \Psi  \right\rangle} } $ and $2j$bit$_{{\left| \Psi  \right\rangle} } $ in the denominator of $\nu _{bit} $ and $\nu _{bit} \left|_{\beta }^{\alpha } \right. $ in (5.9), represent $1{\rm q}$-bit and $1j{\rm q}$-bit, respectively, where index, ${\left| \Psi  \right\rangle} $, denotes any entangled Bell state for binary storable value, `bit' of information. By means of which, $2$bit$_{{\left| \Psi  \right\rangle} } $ or $2j$bit$_{{\left| \Psi  \right\rangle} } $ for signal $B_{in} $, is defined in some Hilbert space $H_{\mathbf{S}} $ of (5.5) to (5.9), for any involved system satisfying ${\left| \Psi  \right\rangle} $ in terms of one-dimensional to N-dimensional expandability; discussed via relations (1.1 to 1.9), Theorem 1.1, and Proof 1.1, \S\ref{section1}. Notation $\mathbf{S}$ in $H_{\mathbf{S}} $, represents the signal subsystem as a state of quantum system, in this case, the storable $B_{in} $ signal. The Hilbert's subsystem time-domain integral on time phase $\phi _{2} $, satisfies the storable signal $B_{in} $ in (5.8) by the system's EDL characteristics, whilst classical signal $A_{in} $ is being stored via bit-frequency $\nu _{bit} $ of (5.9).

In the context of SR, Hilbert space $H_{\mathbf{S}} $'s patterns on quasi-particle as dual carriers in discrete subspaces on carrier $e$'s space would basically derive Tab.$\,$4.1, \S\ref{section1}. The space given to quantum charge as $e$'s space, is described to be a superspace constructed by fields from \S\ref{section4}. Tab.$\,$4.1's consequent equations, resemble with dual superboson (superfermion) characteristics based on quasi-particle algebras, relations (3.3) to (3.5), \S\ref{section3}, by Temme in~\cite{37-Temme}, with the definition of supersymmetric systems, by Popowicz in~\cite{29-Popowicz}, respectively.

\markboth{}{Conclusion and future remarks}
\section{Conclusion and future remarks}
\markboth{}{Conclusion and future remarks}
\vspace{4pt}
\label{section6}
\small The empirical rudimentary concept of data output on the functions of SHAM's architecture from \S\ref{section3}, shall be submitted in the future reports. In those progressing reports, e.g., transmission electron microscopy (TEM imaging) on the chip's components such as CCNTs relevant to the findings of Biró \textit{et al.}~\cite{03-Biro}; Pack \textit{et al.}~\cite{28-Pack}, is concluded once the testability features of the entire components are gathered upon. Unambiguously, in those reports, as exemplified in, \S\ref{section3}, the data collection would be comprised of: a set of phases of data analysis, error reporting procedure and experimentation prior to design objectives and logic. The analysis and presentation stage of data will be based on the experimental methods of~\cite{17-Kirkup} and simulation kits, e.g., `Fast Henry' inductance analysis tool proposed by Pack \textit{et al.} of~\cite{28-Pack}, 2D and 3D physics-based simulation e.g., ATLAS device simulator~\cite{47-Simucad}.

Treatment of storing data by complex plane's depository vector system defining storable qubit(s), permits one to signify the importance of integrating components that detect quantum property of particles in a controlled manner; for instance, briefly, the current paper introduces `Catalysts, Cavity's Quantum Property via CNTs, Memory Cells and CCNTs as CNFETs' in its body map.

The next pace to take is aimed for structured facts once the researcher develops a report on the SHAM chip project. The objective would be a significant step to store data without worrying about the timing process, data's best location in addressing and thereby, on software programming issues, one not being concerned about levels of program coding and compilation(s).

The other implications of this design is accomplishing devices as timepiece and clocks, winding time by themselves, e.g., picking up electromagnetic pulses for radar systems from any satellite and non-satellite application (e.g., planetary phenomena) source, for any communications' wavelength in different geographical locations including our solar system and beyond. The reason would be the physical approach of highly oscillated bits of information acting sufficient in channeling through the blocks of electrical circuit, and then efficiently located and put into secondary memory locations for data-read and write, correspondingly. Paradigms and examples to this account are subjects of the conformed article extending different aspects of this one~\cite{51-Alipour}.

No matter the classical constraints involved in electronics' design and principle, once the grip of this technology holds its grounds by incorporating the \emph{waterfall effect}, circuitry, narrowing down and simplifying data analysers, program processors on the chip, is deemed to be achievable exponentially. For instance,  whilst possessing a magnitude of $\infty $, asymptotically, for the orientation of array displacement as the present issue is organized with, even considering HDD defragmentation process on data would not be necessary based on PTVD-SHAM's technological aspects. There are those algorithms that endure to test results on the chip's compatibility and characteristics specifying,\emph{ frequency narrowbanding of infinite bands of incoming information from the chip's surrounding environment and beyond}.

In conclusion, once this technology is utilized in terms of CCNTs and CNFETs, data computation beyond quantum computational speed is deemed to be achievable. This so-called `breaking the computational speed limit', can happen, if the geometry of the fabrication process is finalized as a freedom of point on circuit's layout, conductivity and responsivity, revealing the expected vales from hidden variables of the experiment.

\begin{flushleft}
\noindent \textbf{Statement of conclusion:} \end{flushleft}
\vspace{4pt}
\noindent --------- \emph{Assume in the laws of quantum physics that, any assigned value as an input is pondered to bring about the notion of `uncertainty'. On the contrary, conjecturing the notion of certain reserved locations for data by pre-defined time in entanglement (predetermined on anticipated time) for uncertain values, is itself upon any individual which tends to read/write data and time, understood as a `certainty'!} ---------

\section*{Acknowledgements}
\markboth{}{Acknowledgments}
\vspace{4pt}
The work described in this paper was supported on the basis of a scientific book and software programming, research project license no: TXU001347562, USA (1998-2007), from the same author. The work was partially carried out at the University \textit{of} Hull, Postgraduate Computing \& Engineering Departments, UK (2006). The author thanks Dr. Habib Alipour \& Dominic S. Zandi \emph{et al.}, for their moral support and basic costs of the project in release. It is highly appreciated, the former senior lecturers, Gerald Goodwin \& Dr. Martin Dickinson~\cite{12-Dickinson} \emph{of} the Faculty \textit{of }Technology, Department \textit{of} Computing, University \textit{of }Lincoln, UK (2005), for their constructive guidelines on, research structuring and computer ethics. All images, the only available table and diagrams, were constructed in software packages such as: `\textit{MS-Word, MS-Visio, WinEdt/MiKTeX, 3D-SMAX, Mathematica}' with other program editors.
\vspace{-2pt}
  \textbf{  }

\end{document}